\documentclass[runningheads]{llncs}

\usepackage{pslatex}
\usepackage{paralist}
\usepackage{amssymb}
\usepackage{amsmath}
\usepackage{mathtools}
\usepackage{color}
\usepackage{stmaryrd}
\usepackage{shuffle}
\usepackage{leftidx}
\usepackage{ifthen}
\usepackage{stackengine}
\usepackage{todonotes}

\newif\ifLongVersion\LongVersionfalse

\ifLongVersion
\usepackage[createShortEnv,conf={normal}]{proof-at-the-end}
\else
\usepackage[createShortEnv,conf={end,no link to proof,restate}]{proof-at-the-end}
\fi

\ifLongVersion
\newenvironment{myLemmaE}{\begin{lemmaE}}{\end{lemmaE}}
\newenvironment{myTextE}{}{}
\else

\fi

\renewcommand{\paragraph}[1]{\vspace*{.2\baselineskip}\noindent{\textbf{#1}}}

\newcommand{\emptyseq}{\varepsilon}
\newcommand{\nat}{\mathbb{N}}
\newcommand{\arity}{\#}
\newcommand{\arityof}[1]{\arity{#1}}

\newcommand{\lenof}[1]{\mathrm{len}({#1})}
\newcommand{\cardof}[1]{\mathrm{card}({#1})}

\newcommand{\width}[1]{\mathrm{wd}({#1})}

\newcommand{\univ}{\mathsf{U}}

\newcommand{\vars}{\mathcal{V}}
\newcommand{\Vars}{\mathbb{V}}

\newcommand{\preds}{\mathbb{P}}
\newcommand{\isdef}{\stackrel{\scalebox{.4}{$\mathsf{def}$}}{=}}
\newcommand{\iffdef}{\stackrel{\hspace*{1.5pt}\scalebox{.4}{$\mathsf{def}$}}{\iff}}
\newcommand{\interv}[2]{[{#1}..{#2}]}
\newcommand{\tuple}[1]{\langle {#1} \rangle}

\newcommand{\set}[1]{\{ {#1} \}}

\newcommand{\pow}[1]{\mathrm{pow}({#1})}

\newcommand{\np}{$\mathsf{NP}$}

\newcommand{\relations}{\mathbb{R}}

\newcommand{\alphabet}{\mathbb{A}}
\newcommand{\betabet}{\mathbb{B}}

\newcommand{\edgrel}[1]{\mathsf{edg}_{#1}}

\newcommand{\arel}{\mathsf{r}}

\newcommand{\acrel}{\mathsf{c}}

\newcommand{\erel}{\mathsf{e}}

\newcommand{\auto}[2]{\mathcal{A}_{
    {#1}
    \ifthenelse{\equal{#2}{}}{}{,{#2}}
}}

\newcommand{\autsat}[2]{\mathcal{A}^{I}_{
    {#1}
    \ifthenelse{\equal{#2}{}}{}{,{#2}}
}}

\newcommand{\autcut}[2]{\mathcal{A}^{\scriptscriptstyle\mathsf{cut}}_{
    {#1}
    \ifthenelse{\equal{#2}{}}{}{,{#2}}
}}

\newcommand{\autcst}[2]{\mathcal{A}^{\scriptscriptstyle\mathsf{cst}}_{
    {#1}
    \ifthenelse{\equal{#2}{}}{}{,{#2}}
  }
}

\makeatletter
\newcommand*{\da@rightarrow}{\mathchar"0\hexnumber@\symAMSa 4B }
\newcommand*{\da@leftarrow}{\mathchar"0\hexnumber@\symAMSa 4C }
\newcommand*{\xdashrightarrow}[2][]{%
  \mathrel{%
    \mathpalette{\da@xarrow{#1}{#2}{}\da@rightarrow{\,}{}}{}%
  }%
}
\newcommand{\xdashleftarrow}[2][]{%
  \mathrel{%
    \mathpalette{\da@xarrow{#1}{#2}\da@leftarrow{}{}{\,}}{}%
  }%
}
\newcommand*{\da@xarrow}[7]{%
  \sbox0{$\ifx#7\scriptstyle\scriptscriptstyle\else\scriptstyle\fi#5#1#6\m@th$}%
  \sbox2{$\ifx#7\scriptstyle\scriptscriptstyle\else\scriptstyle\fi#5#2#6\m@th$}%
  \sbox4{$#7\dabar@\m@th$}%
  \dimen@=\wd0 %
  \ifdim\wd2 >\dimen@
    \dimen@=\wd2 %
  \fi
  \count@=2 %
  \def\da@bars{\dabar@\dabar@}%
  \@whiledim\count@\wd4<\dimen@\do{%
    \advance\count@\@ne
    \expandafter\def\expandafter\da@bars\expandafter{%
      \da@bars
      \dabar@
    }%
  }%
  \mathrel{#3}%
  \mathrel{%
    \mathop{\da@bars}\limits
    \ifx\\#1\\%
    \else
      _{\copy0}%
    \fi
    \ifx\\#2\\%
    \else
      ^{\copy2}%
    \fi
  }%
  \mathrel{#4}%
}
\makeatother

\newcommand{\store}{\mathfrak{s}}
\newcommand{\struc}{\sigma}

\newcommand{\Bigstar}{\mathop{\Asterisk}}

\newcommand{\comp}{\bullet}

\newcommand{\predname}[1]{\mathsf{#1}}

\newcommand{\apred}{\predname{A}}
\newcommand{\bpred}{\predname{B}}
\newcommand{\cpred}{\predname{C}}

\newcommand{\ppred}{\predname{P}}
\newcommand{\qpred}{\predname{Q}}

\newcommand{\emp}{\predname{emp}}

\let\Asterisk\undefined
\newcommand{\Asterisk}{\mathop{\scalebox{1.9}{\raisebox{-0.2ex}{$\ast$}}}\hspace*{1pt}}%
\renewcommand{\vec}[1]{\mathbf #1}
\newcommand{\fv}[1]{\mathrm{fv}({#1})}

\newcommand{\mso}{\textsf{MSO}}
\newcommand{\msone}{$\mathsf{MSO}_1$}

\newcommand{\Models}{\Vdash}

\newcommand{\seplog}{\textsf{SL}}

\newcommand{\slr}{\textsf{SLR}}

\newcommand{\gl}{\textsf{GL}}

\newcommand{\asid}{\Delta}

\newcommand{\arule}{\rho}

\newcommand{\size}[1]{\mathrm{size}({#1})}

\newcommand{\rbr}{{\bf ]\!]}}
\newcommand{\lbr}{{\bf [\![}}
\newcommand{\sem}[1]{{\lbr #1 \rbr}}

\newcommand{\sidsem}[2]{\sem{{#1}}_{#2}}

\newcommand{\csem}[2]{\sem{{#1}}^{\scriptscriptstyle\mathsf{c}}_{#2}}
\newcommand{\rcsem}[2]{\sem{{#1}}^{\scriptscriptstyle\mathsf{r}}_{#2}}

\newcommand{\fusion}[1]{\mathtt{F}^{\scriptscriptstyle {#1}}}
\newcommand{\transfusion}{\mathtt{F}^*}
\newcommand{\fission}[1]{\mathtt{F}^{{\scriptscriptstyle \text{-}{#1}}}}
\newcommand{\transfission}{(\fission{1})^*}

\newcommand{\trparse}{\mathsf{parse}}
\newcommand{\trfission}{\mathsf{fission}}

\newcommand{\sintfusion}[2]{\widetilde{\mathtt{IF}}({#1}\ifthenelse{\equal{#2}{}}{}{,{#2}})}

\newcommand{\diseq}{\mathfrak{d}}

\newcommand{\diseqform}{\mathsf{neq}}

\newcommand{\langof}[2]{\mathcal{L}_{#1}({#2})}

\newcommand{\proj}[2]{{#1}\!\!\downharpoonleft_{\scriptscriptstyle{#2}}}

\newcounter{index}

\newcommand{\eqof}[1]{\approx_{\scriptscriptstyle{#1}}}

\newcommand{\atpos}[2]{{#1}^{\scriptscriptstyle\!\tuple{#2}}}

\newcommand{\funcolor}[1]{\mathit{C}_{#1}}
\newcommand{\fbof}[1]{\mathrm{fb}(#1)}
\newcommand{\nrof}[2]{\sharp_{#1}(#2)}

\newcommand{\graph}{G}
\newcommand{\graphs}{\mathcal{G}}
\newcommand{\eval}[1]{\mathsf{val}({#1})}

\newcommand{\cgraphsof}[1]{\mathcal{G}^c({#1})}
\newcommand{\graphsof}[1]{\mathcal{G}({#1})}
\newcommand{\vertices}{V}
\newcommand{\vertof}[1]{\vertices_{\scriptscriptstyle{#1}}}
\newcommand{\edges}{E}
\newcommand{\edgeof}[1]{\edges_{\scriptscriptstyle{#1}}}
\newcommand{\labels}{\lambda}
\newcommand{\labof}[1]{\labels_{\scriptscriptstyle{#1}}}
\newcommand{\edgerel}{\upsilon}
\newcommand{\edgerelof}[1]{\edgerel_{\scriptscriptstyle{#1}}}

\newcommand{\sources}{\xi}
\newcommand{\sourceof}[1]{\sources_{\scriptscriptstyle{#1}}}

\newcommand{\grammar}{\Gamma}

\newcommand{\trans}{\tau}
\newcommand{\scheme}{\Theta}
\newcommand{\defdof}[2]{\mathrm{def}^{#2}({#1})}
\newcommand{\defd}[1]{\mathrm{def}({#1})}
\newcommand{\hr}{\textsf{HR}}
\newcommand{\rules}{\mathcal{R}}
\newcommand{\nonterm}{\mathcal{U}}

\newcommand{\tree}{T}
\newcommand{\rootof}[1]{\mathsf{root}({#1})}
\newcommand{\reduce}{\leadsto}
\newcommand{\parsetreeof}[2]{\mathcal{T}({#1},{#2})}
\newcommand{\subtree}[2]{{#1}|_{#2}}

\newcommand{\twof}[1]{\mathrm{tw}({#1})}
\newcommand{\charform}[1]{\Theta({#1})}

\newcommand{\funcol}[1]{\mathcal{C}_{#1}}

\begin{document}

\title{Effective MSO-Definability for Tree-width Bounded Models of an
  Inductive Separation Logic of Relations}
\titlerunning{~}

\author{Marius Bozga\inst{1}\orcidID{0000-0003-4412-5684} \and
  Lucas Bueri\inst{1}\orcidID{0000-0002-8589-6955} \and
  Radu Iosif\inst{1}\orcidID{0000-0003-3204-3294}  \and
  Florian Zuleger\inst{2}\orcidID{0000-0003-1468-8398}}
\institute{Univ. Grenoble Alpes, CNRS, Grenoble INP, VERIMAG, 38000, France \and
  Institute of Logic and Computation, Technische Universit\"{a}t Wien, Austria}

\maketitle

\begin{abstract}
  A class of graph languages is definable in Monadic Second-Order
  logic (\mso) if and only if it consists of sets of models of
  \mso\ formul{\ae}.  If, moreover, there is a computable bound on the
  tree-widths of the graphs in each such set, the satisfiability and
  entailment problems are decidable, by Courcelle's
  Theorem~\cite{CourcelleI}. This motivates the comparison of other
  graph logics to \mso. In this paper, we consider the MSO
  definability of a Separation Logic of Relations (\slr) that
  describes simple hyper-graphs, in which each sequence of vertices is
  attached to at most one edge with a given label. Our logic SLR uses
  inductive predicates whose recursive definitions consist of
  existentially quantified separated conjunctions of relation and
  predicate atoms. The main contribution of this paper is an
  expressive fragment of \slr\ that describes bounded tree-width sets
  of graphs which can, moreover, be effectively translated into \mso.
\end{abstract}

\section{Introduction}

Formal languages study finite representations of infinite sets of
objects (e.g., words, trees, graphs). We distinguish between
\emph{descriptive} representations, that define sets by a logical
property of their members (e.g., even numbers, planar graphs, etc.)
and \emph{constructive} representations, explaining how the members of
the set are built. From a system designer's point of view, descriptive
representations are useful for specifying properties (e.g., a shape
invariant of a distributed network), whereas constructive
representations model low-level implementation details (e.g., how the
network components are interconnected). The verification problem
``does each implementation satisfy a given property?'' amounts to
checking if the set of implementations built according to a
constructive representation is included in the descriptive
specification set.

Finite representations of sets of graphs are used in many areas of
computing, e.g., static analysis \cite{10.1145/582153.582161},
databases \cite{AbitebouldBunemanSuciu00} and concurrency
\cite{DBLP:conf/birthday/2008montanari}. A well-established
descriptive representation of graphs is Monadic Second Order Logic
(\mso)~\cite{courcelle_engelfriet_2012}. On the other hand, Hyperedge
Replacement (\hr) grammars are standard constructive representations,
that define sets of graphs as least solutions of a finite set of
recursive equations that uses operations of substitution of a hyperedge
in a graph by another graph~\cite{courcelle_engelfriet_2012}.

The \emph{tree-width} of a graph is a positive integer intuitively
measuring how far the graph is from a tree. For instance, trees have
treewidth one, series-parallel graphs (i.e., circuits with one input
and one output that can be either cascaded or overlaid) have treewidth
two, whereas $n \times n$ square grids have treewidth $n$. This
parameter is important in obtaining efficient algorithms on
graphs. For instance, the \np-complete problems of Hamiltonicity and
3-Coloring can be solved in polynomial time, when restricted to
tree-width bounded classes of graphs ~\cite[Chapter
  11]{DBLP:series/txtcs/FlumG06}.

Bounding the tree-width draws the frontier between the decidability
and undecidability of the satisfiability problem for \mso. Courcelle
\cite{CourcelleI} showed that \mso\ is decidable over bounded
tree-width graphs. Dually, Seese \cite{Seese91} showed that each class
of graphs with a decidable \mso\ theory necessarily has bounded
tree-width. These seminal results motivate finding effective
translations of graph representation languages into \mso. For
instance, Courcelle defined a class of \emph{regular \hr\ graph
  grammars} that describe \mso-definable sets~\cite[\S5]{CourcelleV}
and Doumane developped a language of regular expressions that is
equivalent to the \mso-definable sets of tree-width at most
two~\cite{DBLP:conf/icalp/Doumane22}.

Substructural logics~\cite{Paoli02} are extensions of first order
logic with a non-idempotent \emph{separating conjunction}\footnote{For
  which Gentzen's natural deduction rules of weakening and contraction
  do not hold.}. A distinguished substructural logic is Separation
Logic (\seplog) \cite{Ishtiaq00bias,Reynolds02}, that is interpreted
over finite partial functions of fixed arity, called \emph{heaps}. In
\seplog, the separating conjunction stands for the union of heaps with
disjoint domains. When combined with \emph{inductive
  definitions}~\cite{ACZEL1977739}, \seplog\ gives concise
descriptions of the recursive datastructures (singly- and
doubly-linked lists, trees, etc.) used in imperative programming
(e.g., C, C++, Java, etc.). The shape of these structures can be
described using only existentially quantified separating conjunctions
of (dis-)equalities and points-to atoms. This subset of \seplog\ is
referred to as the \emph{symbolic heap} fragment and is used in
several practical program verification
systems~\cite{CHIN20121006,10.1145/1449764.1449782,10.1007/11804192_6}.

The Separation Logic of Relations (\slr) is a generalization of
\seplog\ to relational structures, first considered for reasoning
about relational databases and object-oriented languages
\cite{10.1007/978-3-540-27864-1_26}. Here, the separating conjunction
splits the interpretation of each relation symbol from the signature
into disjoint parts. For instance, the formula $\arel(x_1, \ldots,
x_n) * \arel(y_1, \ldots, y_n)$ says that $\arel$ is interpreted as a
set with two tuples, i.e., the values of $x_i$ and $y_i$ differ for at
least one index $1 \leq i \leq n$. This style of decomposition is
found in other substructural graph logics, such as the Graph Logic
(\gl) of Cardelli et al \cite{Cardelli2002Spatial}.

Substructural logics constitute powerful tools for reasoning about
mutations of graphs. They can describe actions \emph{locally}, i.e.,
only with respect to the resources (e.g., memory cells, network nodes)
involved, while framing out the part that is irrelevant for that
particular action. This principle of describing mutations, known as
\emph{local reasoning} \cite{CalcagnoOHearnYan07}, is at the heart of
powerful compositional proof techniques for pointer programs
~\cite{CalcagnoDistefanoOHearnYang11}.

\paragraph{Related Work}
Our motivation for studying the symbolic heap-like fragment (i.e.,
formul{\ae} written using inductive predicates whose definitions are
existentially quantified separated conjunctions of relation atoms and
predicates) of \slr\ arose from recent work on deductive verification
of self-adapting distributed systems, where \slr\ is used to write
Hoare-style correctness proofs for component-based and distributed
systems with structured reconfigurable networks
\cite{AhrensBozgaIosifKatoen21,DBLP:conf/concur/BozgaBI22,DBLP:journals/tcs/BozgaIS23}.
A key ingredient of automated proof generation in Hoare logic is the
availability of a decision procedure for the \emph{entailment problem}
$\phi \models_\asid \psi$ asking if each model of a formula $\phi$ is
also a model of another formula $\psi$, when the predicate symbols in
$\phi$ and $\psi$ are interpreted by a set of inductive definitions
$\asid$. We note that, in general, the entailment problem is
undecidable for \slr~\cite[Theorem 3]{DBLP:conf/cade/BozgaBI22}.

One of the first fragments of \seplog\ with a decidable entailment
problem relied on an ad-hoc translation into equivalent
\mso\ formul{\ae}, together with a syntactic guarantee of tree-width
boundedness~\cite{DBLP:conf/cade/IosifRS13}. More recently, this
fragment was the focus of an impressive body of
work~\cite{DBLP:conf/concur/CookHOPW11,DBLP:conf/lpar/KatelaanZ20,EchenimIosifPeltier21b,EchenimIosifPeltier21,DBLP:conf/fossacs/LeL23}.
The entailment problem for \slr\ has been tackled by a reduction to
the same problem for \seplog, that encodes graphs as heaps of fixed
arity and thus works only for graphs of bounded
degree~\cite{DBLP:conf/cade/BozgaBI22}. A goal of the present paper is
to lift the bounded degree restriction and achieve decidability of
entailments in a fragment of \slr, that describes set of graphs of
\emph{unbounded degree}.

Recently, we compared the expressivity of \slr\ with inductive
definitions with that of \mso~\cite{DBLP:conf/concur/IosifZ23}. When
restricting the interpretation of the logics to tree-width bounded
graphs, \slr\ strictly subsumes \mso, i.e., for each \mso\ formula
$\phi$ and integer $k\geq1$, there exists a formula $\psi$ of
\slr\ that defines the models of $\phi$ of tree-width at most
$k$. Moreover, the logics are incomparable for classes of graphs of
unbounded tree-width.

\paragraph{Contribution}
This paper contributes to a dual expressivity result. We define a
fragment of \slr, that can specify rich shape properties of graphs of
unbounded size and degree (e.g., chains, rings, trees, stars, etc.),
whose formul{\ae}~\begin{inparaenum}[(1)]
\item define bounded tree-width sets of graphs, with an effectively
  computable bound, and
\item can be effectively translated into equivalent \mso\ formul{\ae}.
\end{inparaenum}
Our fragment is defined by an easy-to-check condition on the syntax of
the inductive definitions of the predicates used in a formula.

To understand the argument behind the decidability of entailments, let
$\overline\phi$ be the equivalent \mso\ formula obtained from an
\slr\ formula $\phi$ by our translation. Each instance $\phi
\models_\asid \psi$ of the entailment problem reduces to the
unsatisfiablity of the \mso\ formula $\overline\phi \wedge
\neg\overline\psi$. The latter problem is decidable for graphs of
bounded tree-width, by Courcelle's Theorem~\cite[Corollary 4.8
  (2)]{CourcelleI}. In particular, the tree-width of the models of
$\overline\phi\wedge\neg\overline\psi$ is bounded by the property (1)
above and the equivalence of $\phi$ and $\overline\phi$.

For space reasons, the proofs of the results in the paper are given in
Appendix \ref{app:proofs}.

\section{Preliminaries}

We write $\interv{i}{j}$ for the set $\set{i,i+1,\ldots,j}$ of
integers, assumed to be empty if $i>j$. For a set $A$, we denote by
$\pow{A}$ its powerset, $A^0 \isdef \set{\epsilon}$, $A^{i+1} \isdef
A^i \times A$, for all $i \geq 0$, $A^* \isdef \bigcup_{i\geq0} A^i$
and $A^+ \isdef \bigcup_{i\geq1} A^i$, where $\times$ is the Cartesian
product and $\emptyseq$ denotes the empty sequence. Intuitively, $A^*$
(resp. $A^+$) denotes the set of possibly empty (resp. nonempty)
sequences of elements from $A$. The length of a sequence $\alpha \in
A^*$ is denoted by $\lenof{\alpha}$ and its $i$-th element by
$\alpha_i$, for $i \in \interv{1}{\lenof{\alpha}}$.

For a binary relation $R \subseteq A \times B$, we denote by $R^{-1}$
the inverse relation and $R(L)$ for the image of $L$ via $R$, where
$R(a)$ stands for $R(\set{a})$. The composition of relations $R_1
\subseteq A \times B$ and $R_2 \subseteq B \times C$ is denoted as
$R_1 \circ R_2$.  For two sequences of the same length
$\alpha=(a_1,\ldots,a_n) \in A^*$, $\beta=(b_1,\ldots,b_n)\in B^*$ and
a binary relation $R \subseteq A \times B$, we write $\alpha ~R~
\beta$ for $a_i ~R~ b_i$, for all $i \in \interv{1}{n}$.

\begin{figure}[t!]
  \centerline{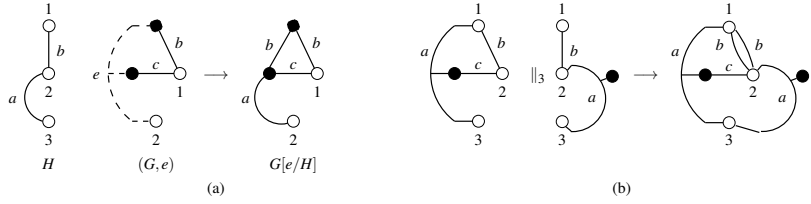}

  \vspace*{-.5\baselineskip}
  \caption{Graph operations: composition (a) parallel composition (b) substitution (c) fusion (d)}
  \label{fig:graphs}
  \vspace*{-.5\baselineskip}
\end{figure}

\paragraph{Graphs}
Let $\alphabet$ be a finite alphabet of edge labels, ranged over by
$a$, equipped with arities $\arityof{a}\geq1$. A \emph{concrete graph
  (c-graph) of type $n\geq0$} is $\graph = \tuple{\vertof{\graph},
  \edgeof{\graph}, \labof{\graph}, \edgerelof{\graph},
  \sourceof{\graph}}$, where: \begin{compactitem}
\item $\vertof{\graph}$ is a finite set of \emph{vertices},
\item $\edgeof{\graph}$ is a finite set of \emph{edges}, such that
  $\vertof{\graph}\cap\edgeof{\graph}=\emptyset$,
\item $\labof{\graph} : \edgeof{\graph} \rightarrow \alphabet$ is a
  mapping that defines the labels of the edges,
\item $\edgerelof{\graph} : \edgeof{\graph} \rightarrow
  \vertof{\graph}^+$ is a mapping that associates each edge a nonempty
  sequence of vertices attached to the edge, such that
  $\arityof{(\labof{\graph}(e))} = \lenof{\edgerelof{\graph}(e)}$, for
  each $e \in \edgeof{\graph}$,
\item $\sourceof{\graph} : \interv{1}{n} \rightarrow \vertof{\graph}$
  is an injective mapping that designates $n$ distinct
  \emph{sources}. The vertex $\sourceof{\graph}(i)$ is the
  \emph{$i$-th source} of $\graph$. A vertex $v \in \vertof{\graph}$
  is \emph{internal} if $v \not\in \sourceof{\graph}(\interv{1}{n})$.  For
  simplicity, we omit specifying sources for graphs of type $0$.
\end{compactitem}
For instance, Fig. \ref{fig:graphs} shows several graphs with internal
vertices drawn as solid circles and sources as shallow circles with
attached numbers. A \emph{path} from $v_1$ to $v_{k+1}$ in $\graph$ is
a sequence of tuples \(\tuple{v_1,i_1,e_1,v_2,j_1}, \ldots,
\tuple{v_k,i_k,e_k,v_{k+1},j_k}\), where $v_1,\ldots,v_{k+1} \in
\vertof{\graph}$, $e_1, \ldots, e_k \in \edgeof{\graph}$,
$v_\ell=\edgerelof{\graph}(e_\ell)_{i_\ell}$ and
$v_{\ell+1}=\edgerelof{\graph}(e_\ell)_{j_\ell}$, for all $\ell \in
\interv{1}{k}$. The path is \emph{rooted} if $i_1 = \ldots i_k = 1$. A
set of vertices $C \subseteq \vertof{\graph}$ is \emph{connected} in
$\graph$ iff for any $u,v\in C$ there exists a path from $u$ to $v$ in
$\graph$ containing only vertices from $C$.

Two c-graphs are \emph{disjoint} iff their sets of vertices and edges
are disjoint, respectively. Two c-graphs are \emph{isomorphic} iff
there exists a bijection between their vertices and edges that
preserves the edge labels, the sequences of vertices attached to edges
and the sources. A \emph{graph} is the equivalence class of a c-graph
for isomorphism. We denote by $\cgraphsof{\alphabet}{}$ the set of
c-graphs with edge labels from $\alphabet$ and by
$\graphsof{\alphabet}$ the set of graphs consisting of c-graphs from
$\cgraphsof{\alphabet}$.

\begin{definition}\label{def:simple-graph}
  A c-graph $\graph = \tuple{\vertof{\graph}, \edgeof{\graph},
    \labof{\graph}, \edgerelof{\graph}, \sourceof{\graph}}$ is
  \emph{simple} iff $\edgerelof{\graph}(e_1) \neq
  \edgerelof{\graph}(e_2)$ for any two edges $e_1,e_2 \in
  \edgeof{\graph}$ with $\labof{\graph}(e_1)=\labof{\graph}(e_2)$. A
  \emph{simple graph} is a graph containing a simple c-graph (the
  class of simple graphs is closed under isomorphism).
\end{definition}
For example, the top-right graph in Fig. \ref{fig:graphs} is not
simple, whereas every other graph in this figure is simple.

The \emph{composition} of c-graphs is a binary operation that produces
a new graph defined as the union of two c-graphs with disjoint sets of
edges:

\begin{definition}\label{def:composition}
  Let $\graph_1$ and $\graph_2$ be two c-graphs of type $0$. The
  c-graphs are \emph{composable} iff~\begin{inparaenum}[(1)]
  \item\label{it1:def:composition} $\edgeof{\graph_1} \cap \edgeof{\graph_2} = \emptyset$,
  \item\label{it2:def:composition} for all $e_i \in \edgeof{\graph_i}$, $i=1,2$, such that
    $\labof{\graph_1}(e_1)=\labof{\graph_2}(e_2)$, we have
    $\edgerelof{\graph_1}(e_1) \neq \edgerelof{\graph_2}(e_2)$.
  \end{inparaenum}
  If $\graph_1$ and $\graph_2$ are composable c-graphs, their
  composition is $\graph_1 \comp \graph_2 \isdef
  \tuple{\vertof{\graph_1} \cup \vertof{\graph_2}, \edgeof{\graph_1}
    \cup \edgeof{\graph_2}, \labof{\graph_1} \cup \labof{\graph_2},
    \edgerelof{\graph_1} \cup \edgerelof{\graph_2}}$, and undefined if
  $\graph_1$ and $\graph_2$ are not composable. The composition of two
  graphs is the graph of the composition of any two composable
  c-graphs taken from each graph.
\end{definition}
In particular, condition (\ref{it2:def:composition}) of
Def. \ref{def:composition} ensures that edges with the same label
attached to the same tuples of vertices prevent the c-graphs from
composing, thus ensuring that the composition is simple c-graphs is a
simple c-graph. For example, Fig. \ref{fig:graphs} (a) shows the
composition of c-graphs with common vertices $u_2$ and $u_3$. Note
that the c-graphs are not composable if $u_6$ is replaced by the $u_1$
in the second c-graph.

\paragraph{Trees}
A \emph{tree} is a c-graph $\tree = \tuple{\vertof{\tree},
  \edgeof{\tree}, \labof{\tree}, \edgerelof{\tree}}$ with a unique
vertex $\rootof{\tree} \in\vertof{\tree}$, where: \begin{compactitem}
\item $\edgerelof{\tree}(e)_1 = \rootof{\tree}$, for each edge $e$
  attached to $\rootof{\tree}$, and
\item there is a single rooted path from $\rootof{\tree}$ to each
  vertex $v \in \vertof{\tree}\setminus\set{\rootof{\tree}}$.
\end{compactitem}
For a vertex $v \in \vertof{\tree}$, we denote by $\subtree{\tree}{v}$
the subtree of $\tree$ having $v$ as root.

We define the \emph{tree decomposition} and \emph{tree-width} of a
c-graph using trees over an alphabet of edge labels having a single
binary element:

\begin{definition}\label{def:tree-decomposition}
  A \emph{tree decomposition} of a c-graph $\graph$ is a pair
  $(\tree,\beta)$, where $\tree$ is a tree and $\beta : \vertof{\tree}
  \rightarrow \pow{\vertof{\graph}}$ maps the vertices of $\tree$ into
  sets of vertices from $\graph$, such that: \begin{compactenum}
  \item\label{it1:treewidth} for each $e \in \edgeof{\graph}$,
    $\edgerelof{\graph}(e) = \tuple{v_1, \ldots, v_k}$, there exists
    $n \in \vertof{\tree}$ such that $v_1, \ldots, v_k \in \beta(n)$,
  \item\label{it2:treewidth} for each $v \in \vertof{\graph}$
    the set $\set{n \in \vertof{\tree} \mid v \in \beta(n)}$ is
    nonempty and connected in $\tree$.
  \end{compactenum}
  The \emph{width} of the tree decomposition is $\width{\tree,\beta}
  \isdef \max_{n \in \vertof{\tree}} \cardof{\beta(n)}-1$. The
  \emph{tree-width} of $\graph$ is $\twof{\graph} \isdef \min
  \set{\width{\tree,\beta} \mid (\tree,\beta) \text{ is a tree
      decomposition of } \graph}$. The tree-width of a graph is the
  tree-width of some c-graph from it\footnote{Isomorphic c-graphs have
  equal tree-widths.}. A set $\graphs$ of graphs has \emph{bounded
  tree-width} iff $\set{\twof{\graph} \mid \graph\in\graphs}$ is a
  finite set.
\end{definition}

\vspace*{-.5\baselineskip}
\paragraph{Graph Grammars}
Let $\graph_1$ and $\graph_2$ be disjoint c-graphs of type $n$. Then
$\graph_1\parallel_n\graph_2$ denotes the c-graph obtained from the
union of $\graph_1$ and $\graph_2$ by joining the pair of $i$-sources
$\tuple{\sourceof{\graph_1}(i), \sourceof{\graph_2}(i)}$, that becomes
the $i$-source of the result, for all $i \in \interv{1}{n}$. The
$\parallel_n$ operation preserves isomorphism, being lifted from
c-graphs to graphs as usual. Fig. \ref{fig:graphs} (b) shows the
$\parallel_3$ operation on two graphs, the result being a non-simple
graph with two $b$-edges between the $1^\mathit{st}$ and
$2^\mathit{nd}$ sources. Note the difference with composition
(Def. \ref{def:composition}) that requires only the edges to be
disjoint. In contrast, $\parallel_n$ keeps both vertices and edges
disjoint, except for the respective $i$-th sources of the operands
that are joined.

For a c-graph $\graph$, an edge $e \in \edgeof{\graph}$ such that
$\arityof{\labof{\graph}}(e)=m$ and a c-graph $H$ of type $m$ disjoint
from $\graph$, the \emph{substitution} $\graph[e/H]$ deletes the edge
$e$ from $\graph$, adds $H$ to $\graph$ and joins the $i$-th vertex
form $\edgerelof{\graph}(e)$ with the $i$-th source of $H$, for all $i
\in \interv{1}{m}$. The sources and type of $\graph[e/H]$ are the same
as for $\graph$. The operation $(\graph, e_1, \ldots, e_k)$, where
$e_1, \ldots, e_k \in \edgeof{\graph}$ are pairwise distinct edges,
takes pairwise distinct c-graphs $H_1, \ldots, H_k$ of types
$\arityof{\labof{\graph}(e_1)}, \ldots,
\arityof{\labof{\graph}(e_k)}$, respectively, and returns the c-graph
$\graph[e_1/H_1, \ldots, e_k/H_k]$, where the substitutions can be
done in any order. This operation preserves isomorphism, being
implicitly lifted from c-graphs to graphs. In this case, the
description of the operation uses a representative of the isomorphism
equivalence class of $(\graph,e_1,\ldots,e_k)$, i.e., $\graph$ is an
c-graph and $e_1,\ldots,e_k$ are edges of $\graph$. For example,
Fig. \ref{fig:graphs} (c) shows the substitution of a type $3$ graph
for a dashed edge $e$ of arity $3$.

\begin{figure}[t!]
  \centerline{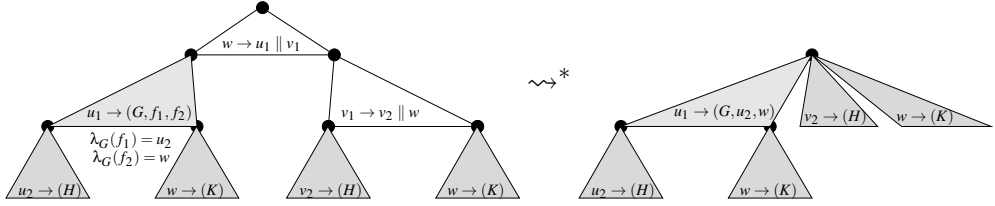}

  \vspace*{-.5\baselineskip}
  \caption{Derivation (left) and parse (right) trees}
  \label{fig:trees}
  \vspace*{-\baselineskip}
\end{figure}

A \emph{graph grammar} is a tuple $\grammar = (\nonterm,\rules)$,
where $\nonterm$ is a finite set of \emph{nonterminals}, ranged over
by $u,v,w$, with arities $\arityof{u}\geq1$ and $\rules$ is a finite
set of rules, either: \begin{compactitem}
\item (unproductive) $u \rightarrow v_1 \parallel_n \ldots \parallel_n
  v_k$, where $\arityof{u}=\arityof{v_1}= \ldots =\arityof{v_k}=n$, or
\item (productive) $u \rightarrow (\graph,f_1,\ldots,f_\ell)$, where
  $\graph\in\cgraphsof{\alphabet\cup\nonterm}$ is a c-graph and $f_1,
  \ldots, f_\ell\in\edgeof{\graph}$ are the edges of $\graph$ having
  labels from $\nonterm$. We say that $f_1, \ldots, f_\ell$ are the
  \emph{nonterminal} edges of $\graph$ and the edges from
  $\edgeof{\graph} \setminus \set{f_1,\ldots,f_\ell}$ are
  \emph{terminal}.
\end{compactitem}
The \emph{parse trees} of a grammar record the partial order in which
the productive rules of the grammar are applied:

\begin{definition}\label{def:grammar-parse-trees}
  Let $\grammar=(\nonterm,\rules)$ be a graph grammar and $w \in
  \nonterm$ be a nonterminal. A \emph{$(\grammar,w)$-derivation tree}
  $U\in \cgraphsof{\rules}$ is a tree whose root is attached to a
  single edge $e$ labeled by a rule from $\rules$ of the form,
  either: \begin{compactitem}
  \item $w \rightarrow u_1 \parallel_n \ldots \parallel_n u_k$ of arity
    $k$ and $\subtree{U}{\edgerelof{U}(e)_i}$ is a
    $(\grammar,u_i)$-derivation tree, $i \in \interv{1}{k}$, or
  \item $w \rightarrow (\graph,f_1,\ldots,f_\ell)$ of arity $\ell$ and
    $\subtree{U}{\edgerelof{U}(e)_i}$ is a $(\grammar,
    \labof{\graph}(f_i))$-derivation tree, $i \in \interv{1}{\ell}$.
  \end{compactitem}
  We write $U \reduce V$ iff $V$ is obtained from $U$ by deleting an
  edge $e\in\edgeof{U}$, labeled by an unproductive rule, and
  attaching each subtree $\subtree{U}{\edgerelof{U}(e)_i}$ to
  $\edgerelof{U}(e)_1$, for all $i \in
  \interv{2}{\arityof{\labof{U}(e)}}$. A \emph{$(\grammar,w)$-parse
  tree} is a tree $\tree$ labeled by productive rules only, such that
  $U \reduce^* \tree$, for some $(\grammar,w)$-derivation tree. We
  denote by $\parsetreeof{\grammar}{w}$ the set of
  $(\grammar,w)$-parse trees.
\end{definition}
Fig. \ref{fig:trees} shows a derivation tree and its corresponding
parse tree, with edges labeled by productive rules drawn as shaded
triangles. In particular, each vertex of a derivation tree has a
bounded (by the maximal number of nonterminals on the right-hand side
of a grammar rule) number of children, unlike the case of parse trees,
that may have unbounded degrees.

For a derivation tree $U$, we denote by $\eval{U}$ the graph obtained
by interpreting the labels of the edges of $U$ as graph operations,
lifted to sets of graphs in the usual way. Formally, if $e$ is the
(unique) edge such that $\rootof{U}=\edgerelof{U}(e)_1$, we define:
\begin{align*}
  \eval{U} \isdef \left\{\begin{array}{ll}
  \eval{\subtree{U}{\edgerelof{U}(e)_1}} \parallel_n
  \ldots \parallel_n \eval{\subtree{U}{\edgerelof{U}(e)_k}}, & \text{if }
  \labof{U}(e)=[w \rightarrow u_1 \parallel_n \ldots \parallel_n u_k]
  \\
  \graph[f_1/\eval{\subtree{U}{\edgerelof{U}(e)_1}}, \ldots,
      f_\ell/\eval{\subtree{U}{\edgerelof{U}(e)_\ell}}], & \text{if }
    \labof{U}(e)=[w\rightarrow (\graph,f_1,\ldots,f_\ell)]
  \end{array}\right.
\end{align*}
Since the evaluation is invariant modulo commutativity and
associativity of $\parallel_n$, we define
$\eval{\tree}\isdef\eval{U}$, where $\tree$ is a parse tree and $U
\reduce^* \tree$ is any derivation tree. The \emph{language of
  $\grammar$ for a nonterminal $w$} is $\langof{w}{\grammar} \isdef
\eval{\parsetreeof{\grammar}{w}}$. A set of graphs $\mathcal{L}$ is
\emph{hyperedge-replacement} (\hr) if
$\mathcal{L}=\langof{w}{\grammar}$ for a grammar $\grammar$ and a
nonterminal $w$.

\section{Logics}

We introduce two logics interpreted over graphs and state the problem
concerning the translation of formul{\ae} from one logic into
equivalent formul{\ae} in the other.

\begin{figure}[t!]
  \vspace*{-1.5\baselineskip}
  {\footnotesize\begin{center}
    \[\psi := x=y \mid \edgrel{a}(x_1, \ldots, x_{\arityof{a}+1}) \mid X(x) \mid
    \neg\psi \mid \psi \wedge \psi \mid \exists x ~.~ \psi \mid \exists X
    ~.~ \psi\]

    \vspace*{-2mm}(a) \mso\ Syntax

    \vspace*{-4mm}
    \[\begin{array}{rclcl}
    \graph & \Models^\store & x=y & \iff & \store(x)=\store(y) \\
    \graph & \Models^\store & \edgrel{a}(x_1, \ldots, x_{\arityof{a}+1}) & \iff &
    \store(x_1) \in \edgeof{\graph},~ \labof{\graph}(\store(x_1))=a \text{ and }
    \edgerelof{\graph}(\store(x_1))=\tuple{\store(x_2), \ldots, \store(x_{\arityof{a}+1})} \\
    \graph & \Models^\store & X(x) & \iff & \store(x) \in \store(X) \\
    \graph & \Models^\store & \exists X ~.~ \psi & \iff & \graph \Models^{\store[X\leftarrow U]} \psi
    \text{, for a set } U \subseteq \vertof{\graph} \cup \edgeof{\graph}
    \end{array}\]

    \vspace*{-2mm}(b) \mso\ Semantics

    \vspace*{-4mm}
    \[\phi := \emp \mid x=y \mid x\neq y \mid a(x_1, \ldots, x_{\arityof{a}}) \mid \apred(x_1, \ldots,
    x_{\arityof{\apred}}) \mid \phi * \phi \mid \exists x ~.~ \phi\]

    \vspace*{-2mm}(c) \slr\ Syntax

    \vspace*{-4mm}
     \[\begin{array}{rclcl}
     \graph & \models^\store_\asid & \emp & \iffdef & \vertof{\graph}
     = \edgeof{\graph} = \emptyset \\ \graph & \models^\store_\asid &
     x \sim y & \iffdef & \graph \models^\store_\asid \emp \text{ and
     } \store(x) \sim \store(y) \text{, for } \sim \ \in\!\set{=,\neq}
     \\
     \graph & \models^\store_\asid & a(x_1, \ldots,
     x_{\arityof{a}}) & \iffdef & \edgeof{\graph} =
     \set{e},~ \vertof{\graph} = \set{\store(x_1), \ldots,
       \store(x_{\arityof{a}})},~ \labof{\graph}(e) = a
     \text{ and } \edgerelof{\graph}(e) =
     \tuple{\store(x_1), \ldots, \store(x_{\arityof{a}})}, \\
     \graph & \models^\store_\asid & \apred(y_1, \ldots, y_{\arityof{\apred}}) & \iffdef &
     \graph \models^\store_\asid \phi[x_1/y_1, \ldots, x_{\arityof{\apred}}/y_{\arityof{\apred}}]
     \text{, for some rule } \apred(x_1, \ldots, x_{\arityof{\apred}}) \leftarrow \phi \text{ from } \asid
     \\
     \graph & \models^\store_\asid & \phi_1 * \phi_2 &
     \iffdef & \text{there exist graphs } \graph_1,~\graph_2 \text{, such
       that } \graph = \graph_1 \comp \graph_2 \text{ and } \graph_i
     \models^\store_\asid \phi_i, \text{ for } i = 1,2
     \\
     \graph & \models^\store_\asid & \exists x ~.~ \phi & \iffdef & \struc
     \models^{\store[x\leftarrow u]}_\asid \phi \text{, for some
       vertex } u \in \vertof{\graph}
     \end{array}\]

     \vspace*{-2mm}(d) \slr\ Semantics
  \end{center}}
  \vspace*{-\baselineskip}
  \caption{Logics}
  \label{fig:logics}
  \vspace*{-\baselineskip}
\end{figure}

\paragraph{Monadic Second Order Logic} (\mso) is the set of formul{\ae}
written using a set $\vars = \set{x,y,\ldots}$ of \emph{first-order
variables}, a set $\Vars = \set{X,Y,\ldots}$ of \emph{second-order
variables} and relation symbols $\edgrel{a}$ of arity
$\arityof{\edgrel{a}} = \arityof{a}+1$, for all $a \in \alphabet$,
using the syntax in Fig. \ref{fig:logics} (a). A variable is
\emph{free} if it occurs outside the scope of any quantifier. The set
of free variables of a formula $\phi$ is written $\fv{\phi}$. A
\emph{sentence} is a formula without free variables.

The semantics of \mso\ is given by a satisfaction relation $\graph
\Models^\store \psi$ between c-graphs and formul{\ae}, where the store
$\store : \vars \cup \Vars \rightarrow
\vertof{\graph}\cup\edgeof{\graph} \cup
\pow{\vertof{\graph}\cup\edgeof{\graph}}$ maps each variable $x \in
\vars$ to a vertex or an edge and each variable $X \in \Vars$ to a
finite subset of vertices and edges from $\graph$. Because
quantification is over finite sets, the version of \mso\ considered
here is known as \emph{weak} \mso\ in the literature.

The satisfaction relation is defined inductively on the structure of
formul{\ae} in Fig. \ref{fig:logics} (b). The semantics of negation,
conjunction and first-order quantification are standard and
omitted. If $\phi$ is a sentence, the satisfaction relation does not
depend on the store and we write $\graph \Models \phi$ instead of
$\graph \Models^\store \phi$. A set $\graphs$ of c-graphs is
\mso-definable iff there exists a \mso\ sentence $\phi$ such that
$\graphs = \set{\graph \mid \graph \Models \phi}$. It is
known~\cite{DBLP:books/daglib/0082516} that \mso-definable sets are
unions of graphs (i.e., equivalence classes of isomorphic c-graphs),
hence we interpret \mso\ formul{\ae} over graphs in a transparent
fashion.

\paragraph{Separation Logic of Relations} (\slr) uses a set of
\emph{predicates} $\preds = \set{\apred, \bpred, \ldots}$ with given
arities $\arityof{\apred}\geq0$. A predicate of zero arity is called
\emph{nullary}. To alleviate notation, we denote by $\apred$ the
predicate atom $\apred()$, whenever $\apred$ is nullary. The
formul{\ae} of \slr\ are defined by the syntax in
Fig. \ref{fig:logics} (c). Instead of the boolean conjunction,
\slr\ has a \emph{separating conjunction} $*$. The formul{\ae} $x\neq
y$ and $\apred(x_1,\ldots,x_{\arityof{\apred}})$ are called
\emph{disequalities} and \emph{predicate atoms}, respectively. A
formula without predicate atoms is called \emph{predicate-free}. A
\emph{qpf} formula is both quantifier- and predicate-free.

\begin{definition}\label{def:sid}
  A \emph{set of inductive definitions (SID)} is a \emph{finite} set
  of \emph{rules} of the form $\apred(x_1, \ldots,
  x_{\arityof{\apred}}) \leftarrow \phi$, where $x_1, \ldots,
  x_{\arityof{\apred}}$ are pairwise distinct variables, called
  \emph{parameters} and $\phi$ is a \slr\ formula, such that
  $\fv{\phi} \subseteq \set{x_1, \ldots, x_{\arityof{\apred}}}$. We
  say that the rule \emph{defines} $\apred$.
\end{definition}
The semantics of \slr\ is given by the satisfaction relation $\graph
\models^\store_\asid \phi$ between c-graphs and formul{\ae},
parameterized by a store $\store$ and a SID $\asid$.  We write
$\store[x \leftarrow u]$ for the store that maps $x$ into $u$ and
agrees with $\store$ on all variables other than $x$.  By $[x_1/y_1,
  \ldots, x_n/y_n]$ we denote the substitution that replaces each free
variable $x_i$ by $y_i$ in a formula $\phi$, the result of applying
the substitution being denoted as $\phi[x_1/y_1, \ldots, x_n/y_n]$. As
a convention, the existentially quantified variables from $\phi$ are
renamed to avoid clashes with $y_1, \ldots, y_n$. Then
$\models^\store_\asid$ is the least relation that satisfies the
constraints in Fig. \ref{fig:logics} (d).

\begin{figure}[t!]
  \centerline{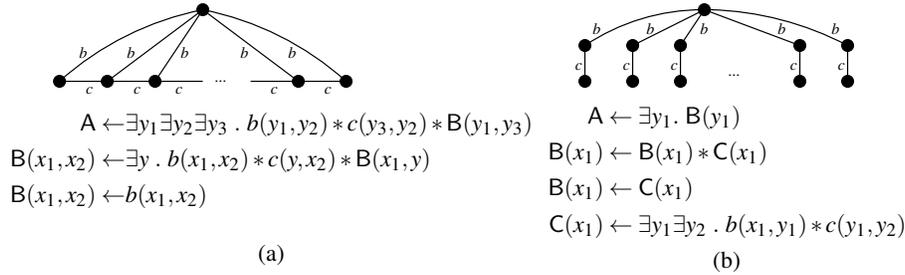}

  \vspace*{-\baselineskip}
  \begin{center}
    \begin{minipage}{.59\textwidth}
      {\footnotesize\begin{align*}
        \apred \leftarrow & \exists y_1 \exists y_2 \exists y_3 ~.~ b(y_1,y_2) * c(y_3,y_2) * \bpred(y_1, y_3) \\
        \bpred(x_1, x_2) \leftarrow & \exists y ~.~ b(x_1,x_2) * c(y,x_2) * \bpred(x_1, y) \\
        \bpred(x_1, x_2) \leftarrow & b(x_1, x_2) \\
        \end{align*}

        \vspace*{-\baselineskip}
        \centerline{(a)}}
    \end{minipage}
    \begin{minipage}{.39\textwidth}
      {\footnotesize\begin{align*}
        \apred & \leftarrow \exists y_1.~ \bpred(y_1) \\
        \bpred(x_1) & \leftarrow \bpred(x_1) * \cpred(x_1) \\
        \bpred(x_1) & \leftarrow \cpred(x_1) \\
        \cpred(x_1) & \leftarrow \exists y_1 \exists y_2 ~.~ b(x_1,y_1) * c(y_1, y_2)
        \end{align*}

        \vspace*{-.5\baselineskip}
        \centerline{(b)}}
    \end{minipage}
  \end{center}

  \vspace*{-1.5\baselineskip}
  \caption{Regular SIDs with productive only (a) and both productive and unproductive (b) rules}
  \label{fig:sids}
  \vspace*{-\baselineskip}
\end{figure}

Note that the interpretation of equalities and relation atoms differs
in \slr\ from \mso. Namely, in \slr, $x = y$ requires that the c-graph
is empty and $a(x_1,\ldots,x_{\arityof{a}})$ denotes the simple c-graph
consisting of one edge attached to the vertices that are the store
values of $x_1, \ldots, x_{\arityof{a}}$, in this order. In contrast
with \mso, \slr\ cannot directly refer to edges by (first-order)
variables.

If $\phi$ is a sentence (resp. a predicate-free formula), we omit the
store $\store$ (resp. the SID $\asid$) from $\graph
\models^\store_\asid \phi$. We write $\sidsem{\phi}{\asid} \isdef
\set{\graph \mid \graph \models_\asid \phi}$ whenever $\phi$ is a
sentence. A sentence $\phi$ is \emph{satisfiable for $\asid$} iff
$\sidsem{\phi}{\asid} \neq \emptyset$. If $\phi$ is predicate-free we
write $\sidsem{\phi}{}$ instead of $\sidsem{\phi}{\asid}$. In this
case, the satisfiability of $\phi$ is defined without reference to a
particular SID. For example, Fig. \ref{fig:sids} shows the sets of
models for two SIDs. These sets contain graphs of unbounded degree,
where the top vertex is connected to an unbounded number of edges.

A set $\graphs$ of c-graphs is \emph{\slr-definable} iff there exists
a SID $\asid$ and a sentence $\phi$ such that $\graphs =
\sidsem{\phi}{\asid}$. It has been shown that \slr-definable sets are
unions of graphs~\cite[Proposition
  7]{DBLP:conf/concur/IosifZ23}. Hence, we shall use \slr\ formul{\ae}
for graphs in a transparent fashion.

\paragraph{\mso-definability of \slr}
We consider the problem of finding, for each \slr\ sentence $\phi$ and
SID $\asid$, an \mso\ formula that defines
$\sidsem{\phi}{\asid}$. However, this problem is undecidable:

\begin{propositionE}[][category=proofs]
  The following is undecidable: given an SID $\asid$ and an
  \slr\ sentence $\phi$, does there exist an \mso\ sentence $\psi$
  such that $\graph \models_\asid \phi \iff \graph \Models \psi$, for
  all graphs $\graph$?
\end{propositionE}
\begin{proofE}
  Let $w = a_1 \ldots a_n \in \set{0,1}^*$ be a finite word. We
  encode $w$ by a c-graph $\graph_w$ of type $0$,
  where: \begin{compactitem}
  \item $\vertof{\graph_w} = \set{v_1,\ldots,v_{n}}$,
  \item $\edgeof{\graph_w} = \set{d_1,\dots,d_n} \cup
    \set{e_1,\ldots,e_{n-1}} \cup \set{f_1,\ldots,f_n}$,
  \item $\labof{\graph_w}(d_i) = \mathfrak{D}$, for all
    $i\in\interv{1}{n}$, $\labof{\graph_w}(e_i) = \mathfrak{S}$, for
    all $i \in \interv{1}{n-1}$ and $\labof{\graph_w}(f_i) = a_i \in
    \set{0,1}$, for all $i \in \interv{1}{n}$,
  \item $\edgerelof{\graph_w}(d_i) = \tuple{v_i}$, for all
    $i\in\interv{1}{n}$, $\edgerelof{\graph_w}(e_i) =
    \tuple{v_i,v_{i+1}}$, for all $i \in \interv{1}{n-1}$ and
    $\edgerelof{\graph_w}(f_i) = \tuple{v_i}$, for all $i \in
    \interv{1}{n}$.
  \end{compactitem}
  where $0,1$ and $\mathfrak{D}$ are unary edge labels and
  $\mathfrak{S}$ is a binary edge label, i.e., the edge label alphabet
  of $\graph_w$ is $0,1,\mathfrak{D},\mathfrak{S}$.

  A \emph{context-free} word grammar $\Gamma = (\mathcal{N},\mathcal{P})$ consists of a
  set of nonterminals $\mathcal{N}$ and a set of productions $\mathcal{P}$ of the form $N
  \rightarrow w$, where $w \in (\mathcal{N} \cup \set{0,1})^*$ is a word. A
  context-free language is a set of words $w \in \set{0,1}^*$ produced
  by a context-free word grammar via derivations $N \longrightarrow^*
  w$, where a derivation step $u \longrightarrow v$ replaces a
  nonterminal from $u$ with the right-hand side of a production. The
  following problem is undecidable~\cite{Greibach68,Bchi1960WeakSA}:
  given a context-free language $\mathcal{L}$, is there a
  \mso\ sentence $\phi$ such that $w\in\mathcal{L} \iff \graph_w
  \Models \phi$. This is consequence of two independent results: the
  problem $\mathcal{L}$ is recognizable is undecidable
  \cite{Greibach68} and $\mathcal{L}$ is recognizable iff $\graphs
  \isdef \set{\graph_w \mid w \in \mathcal{L}}$ is \mso-definable
  \cite{Bchi1960WeakSA}.

  To prove undecidability of the problem in the statement, we encode
  an arbitrary context-free word grammar $\Gamma =
  (\mathcal{N},\mathcal{P})$ by an SID $\asid$, as
  follows: \begin{compactitem}
  \item for each nonterminal $N \in \mathcal{N}$ we have two predicates: $N_0$
    nullary and $N_2$ binary and a rule in $\asid$:
    \begin{align*}
      N_0 \leftarrow \exists y_1 \exists y_2 ~.~ N_2(y_1,y_2)
    \end{align*}
  \item for each production $N \rightarrow \alpha_1 \ldots \alpha_n
    \in \mathcal{P}$, we consider the following rule in $\asid$:
    \begin{align*}
      N_2(x_1,x_2) \leftarrow & \exists y_1 \ldots \exists y_{n+1} ~.~
      x_1 = y_1 * x_2 = y_{n+1} * \Bigstar_{j=1}^{n} \beta_j \\
      \text{ where } \beta_j \isdef & \left\{\begin{array}{ll}
      a_i(y_j) * \mathfrak{D}(y_j) * \mathfrak{S}(y_j,y_{j+1}) \text{, if } \alpha_j = a_i \in \set{0,1} \\
      M_2(y_j,y_{j+1}) \text{, if } \alpha_j = M \in \mathcal{N}
      \end{array}\right.
    \end{align*}
  \end{compactitem}
  Let $\mathcal{L}$ be the language of $\Gamma$ starting with $N$.
  The proof is concluded by checking that $w \in \mathcal{L} \iff
  \graph_w \models_\asid N$ and taking the predicate atom $N$ as
  $\phi$ in the statement of the Proposition. \qed
\end{proofE}

\noindent This negative result motivates the search for a fragment of
\slr\ that is expressive enough to describe interesting shape
properties (e.g., lists, rings, trees, stars and beyond) and for which
an effective translation to \mso\ is possible.

\section{An \mso-Definable Fragment of \slr}
\label{sec:mso-def}

For simplicity, in the rest of this paper we shall represent sentences
$\phi$ by nullary predicate atoms $\apred$. This loses no generality
since $\sidsem{\phi}{\asid} = \sidsem{\apred}{\asid \cup \set{\apred
    \leftarrow \phi}}$ provided that $\apred$ is not defined by any
other rule in $\asid$. We give a first syntactic restriction of the
rules of $\asid$, inspired from the definition of \emph{regular graph
  grammars} introduced by Courcelle~\cite{CourcelleV}. For
self-completness, we recall this definition here, under a slightly
different notation:

\begin{definition}[\cite{CourcelleV}]\label{def:regular-graph-grammar}
  A graph operation $(\graph, f_1,\ldots,f_k)$ is \emph{regular} iff
  the following hold: \begin{compactenum}
  \item\label{it1:regular-graph-operation} $\graph$ has at least one
    edge, either \begin{inparaenum}[(a)]
  \item\label{it11:regular-graph-operation} a single terminal edge
    attached to sources only, or
  \item\label{it12:regular-graph-operation} each of its edges is
    attached to an internal vertex, and
  \end{inparaenum}
  \item\label{it2:regular-graph-operation} between any two vertices of
    $\graph$ there is a path that traverses only internal vertices
    (excepting the endpoints) and terminal edges.
  \end{compactenum}
  A graph grammar $\grammar = (\nonterm,\rules)$ is \emph{regular} iff
  there exists a set $\mathcal{W} \subseteq \nonterm$ of nonterminals
  and the rules in $\rules$ are of one of the following forms, either:
  \begin{compactenum}[A.]
  \item\label{it1:regular-graph-grammar} $w \rightarrow
    (\graph,f_1,\ldots,f_k)$, for some nonterminal $w\in\mathcal{W}$
    and $\labof{G}(f_1),\ldots,\labof{G}(f_k) \in \nonterm$, where
    $(\graph,f_1,\ldots,f_k)$ is a regular graph operation,
  \item\label{it2:regular-graph-grammar} $u \rightarrow u
    \parallel_n w$, where $u \in \nonterm\setminus\mathcal{W}$
    such that $\arityof{u}=n$ and $w \in \mathcal{W}$,
  \item\label{it3:regular-graph-grammar} $u \rightarrow w_1
    \parallel_n \ldots \parallel_n w_k$, $u \in
    \nonterm\setminus\mathcal{W}$ such that $\arityof{u}=n$ and
    $w_1,\ldots,w_k\in\mathcal{W}$.
  \end{compactenum}
\end{definition}
It has been proved that the set of graphs $\langof{w}{\grammar}$ is
\mso-definable, for any nonterminal $w$, if the grammar $\grammar$ is
regular \cite[Theorems 4.8 and 5.10]{CourcelleV}.

We translate rule-by-rule any \hr\ grammar into an SID, based on the
following intuition. The nonterminals of the grammar correspond to the
predicates of the SID. Any productive grammar rule $w \rightarrow
(\graph,f_1,\ldots,f_k)$, where $\graph$ is a c-graph of type $n$,
corresponds to a SID rule:
\vspace*{-\baselineskip}
\begin{align}
  \qpred(x_1,\ldots,x_n) \leftarrow \exists y_1 \ldots \exists y_n ~.~
  \psi * \Bigstar_{j=1}^k \ppred_j(z_{j,1},\ldots,z_{j,\arityof{\ppred_j}})
  \label{rule:prod}
\end{align}

\vspace*{-\baselineskip}\noindent
where $\qpred$ corresponds to $w$, the parameters
$x_1,\ldots,x_{\arityof{\qpred}}$ correspond to the sources of
$\graph$, $\psi$ is a qpf formula consisting of relation atoms
$a(z_1,\ldots,z_{\arityof{a}})$ that correspond to the terminal edges
with label $a$ of $\graph$ and
$\ppred_j(z_{j,1},\ldots,z_{j,\arityof{\ppred_j}})$ are the
nonterminal edges with label $\ppred_j$, corresponding to
$\labof{\graph}(f_j)$, for all $j \in \interv{1}{k}$. With these
conventions, the conditions of Def. \ref{def:regular-graph-grammar}
are transposed from graph grammars to SIDs, as follows:

\begin{definition}\label{def:regular-sid}
  A SID $\asid$ is \emph{regular} iff there exists a set $\mathbb{Q}
  \subseteq \preds$ of \emph{productive predicates}, such that the
  rules in $\asid$ are of the following forms,
  either: \begin{compactenum}
  \item\label{it1:def:regular-sid} (\emph{productive})
    $\qpred(x_1,\ldots,x_{\arityof{\qpred}}) \leftarrow a(z_1, \ldots,
    z_{\arityof{a}})$, where $\qpred\in\mathbb{Q}$; this case
    corresponds to grammar rules of the form
    (\ref{it1:regular-graph-grammar}), with graph operation
    (\ref{it11:regular-graph-operation}) in
    Def. \ref{def:regular-graph-grammar},
  \item\label{it2:def:regular-sid} (\emph{productive})
    $\qpred(x_1,\ldots,x_{\arityof{\qpred}}) \leftarrow \exists y_1
    \ldots \exists y_m ~.~ \psi * \Bigstar_{i=1}^k
    \ppred_i(z_{i,1}, \ldots, z_{i, \arityof{\ppred_i}})$,
    where $\qpred\in\mathbb{Q}$ and $\psi$ is a qpf formula such that
    the following hold: \begin{compactenum}
    \item\label{it21:def:regular-sid} $\{z_{i,1}, \ldots, z_{i,
      \arityof{\ppred_i}}\} \cap \{y_1,\ldots,y_m\} \neq \emptyset$,
      for all $i \in \interv{1}{k}$; see
      Def. \ref{def:regular-graph-grammar}
      (\ref{it12:regular-graph-operation}),
    \item\label{it22:def:regular-sid} for each two variables $z \neq z' \in
      \set{x_1,\ldots,x_{\arityof{\qpred}}} \cup \set{y_1,\ldots,y_m}$
      there exists a sequence of relation atoms
      $a_1(\xi_{1,1},\ldots,\xi_{1,\arityof{a_1}})$, $\ldots$,
      $a_h(\xi_{h,1},\ldots,\xi_{h,\arityof{a_h}})$ occurring
      in $\psi$ such that $z \in
      \set{\xi_{1,1},\ldots,\xi_{1,\arityof{a_1}}}$, $z' \in
      \set{\xi_{h,1},\ldots,\xi_{h,\arityof{a_h}}}$ and
      $\set{\xi_{j,1},\ldots,\xi_{j,\arityof{a_j}}} \cap
      \set{\xi_{j+1,1},\ldots,\xi_{{j+1},\arityof{a_{j+1}}}} \cap
      \set{y_1,\ldots,y_m} \neq \emptyset$, for all $j \in
      \interv{1}{h-1}$; see Def. \ref{def:regular-graph-grammar}
      (\ref{it2:regular-graph-operation}).
    \end{compactenum}
    This case corresponds to grammar rules
    Def. \ref{def:regular-graph-grammar}
    (\ref{it1:regular-graph-grammar}), with graph operation
    (\ref{it12:regular-graph-operation}).
  \item\label{it3:def:regular-sid} (\emph{unproductive})
    $\ppred(x_1,\ldots,x_{\arityof{\ppred}}) \leftarrow
    \ppred(x_{1},\ldots,x_{\arityof{\ppred}}) *
    \qpred(x_{1},\ldots,x_{\arityof{\qpred}})$, where
    $\ppred\in\preds\setminus\mathbb{Q}$ and $\qpred\in\mathbb{Q}$,
    such that $\arityof{\ppred}=\arityof{\qpred}$; this case
    corresponds to grammar rules Def. \ref{def:regular-graph-grammar}
    (\ref{it2:regular-graph-grammar}),
  \item\label{it4:def:regular-sid} (\emph{unproductive})
    $\ppred(x_1,\ldots,x_{\arityof{\ppred}}) \leftarrow
    \Bigstar_{i=1}^\ell\qpred_i(x_1,\ldots,x_{\arityof{\qpred_i}})$,
    where $\ppred\in\preds\setminus\mathbb{Q}$,
    $\qpred_i\in\mathbb{Q}$ and $\arityof{\ppred}=\arityof{\qpred_i}$,
    for all $i \in \interv{1}{\ell}$; this case corresponds to grammar
    rules Def. \ref{def:regular-graph-grammar}
    (\ref{it3:regular-graph-grammar}).
  \end{compactenum}
\end{definition}
For example, the SIDs from Fig. \ref{fig:sids} are both regular,
Fig. \ref{fig:sids} (a) has only productive rules (with
$\mathbb{Q}=\set{\apred,\bpred}$ and
$\preds\setminus\mathbb{Q}=\emptyset$), and Fig. \ref{fig:sids} (b)
has both productive (with $\mathbb{Q}=\set{\apred,\cpred}$) and
unproductive (with $\preds\setminus\mathbb{Q}=\set{\bpred}$) rules.

Similar to graph grammars (Def. \ref{def:grammar-parse-trees}), a
\emph{parse tree} for an SID records the partial order of the
productive rules of the SID that define the satisfaction relation for
a given \slr\ formula. To define parse trees formally, we view each
productive rule of the form (\ref{rule:prod}), where $\qpred \in
\mathbb{Q}$ and $\psi$ is a qpf formula, as a tree edge label of arity
$k+1$.

\begin{definition}\label{def:parse-tree}
  Let $\ppred \in \preds$ be a predicate. A
  \emph{$(\asid,\ppred)$-parse tree} is a tree $\tree$, such that,
  letting $e_1,\ldots,e_m$ be the edges attached to
  $\rootof{\tree}$, the following hold: \begin{compactenum}
  \item\label{it11:def:parse-tree} if $\ppred \in \mathbb{Q}$ then
    $m=1$ and $\labof{\tree}(e_1)$ is a productive rule that defines $\ppred$,
  \item\label{it12:def:parse-tree} else (i.e., $\ppred \in \mathbb{P}
    \setminus \mathbb{Q}$), there exists $\ell\in\interv{0}{m}$ such
    that: \begin{compactenum}
      \item\label{it121:def:parse-tree} there exists a rule
        \(\ppred(x_1,\ldots,x_{\arityof{\ppred}}) \leftarrow
        \Bigstar_{i=1}^\ell\qpred_i(x_1,\ldots,x_{\arityof{\qpred_i}})\),
        such that $\qpred_i \in \mathbb{Q}$ and $\labof{\tree}(e_i)$
        is a productive rule that defines $\qpred_i$, for all $i \in
        \interv{1}{\ell}$,
      \item\label{it122:def:parse-tree} for each $j \in
        \interv{\ell+1}{m}$ there exists a rule
        \(\ppred(x_1,\ldots,x_{\arityof{\ppred}}) \leftarrow
        \ppred(x_{1},\ldots,x_{\arityof{\ppred}}) *
        \qpred(x_{1},\ldots,x_{\arityof{\qpred}})\), such that $\qpred
        \in \mathbb{Q}$ and $\labof{\tree}(e_j)$ is a rule that
        defines $\qpred$.
    \end{compactenum}
  \item\label{it2:def:parse-tree} for each edge $e_i$, letting
    $\labof{\tree}(e_i)$ be a rule of the form (\ref{rule:prod}), each vertex
    $\edgerelof{\tree}(e_i)_{j+1}$ is the root of a
    $(\asid,\ppred_j)$-parse tree, for all $i \in \interv{1}{m}$ and
    $j \in \interv{1}{\arityof{\labof{\tree}(e_i)}}$.
  \end{compactenum}
  We denote by $\parsetreeof{\asid}{\ppred}$ the set of
  $(\asid,\ppred)$-parse trees.
\end{definition}
From each parse tree $\tree$ we build a \emph{characteristic
  formula} $\charform{\tree}$ that describes those $\asid$-models of
$\apred$ derived according to $\tree$. For technical reasons related
to bookkeeping, each free variable $x \in \fv{\charform{\tree}}$ is
annotated with the edge $e\in\edgeof{\tree}$ that introduced the
variable, as in $\atpos{x}{e}$. For a qpf formula $\psi$, we denote by
$\atpos{\psi}{e}$ the formula obtained by annotating each free
variable of $\psi$ with $e$ in this way. Formally, the characteristic
formula of a $(\asid,\ppred)$-parse tree $\tree$ is defined
recursively as:
\vspace*{-\baselineskip}
\begin{align}
  \charform{\tree} \isdef \Bigstar_{i=1}^m
  \Bigstar_{j=1}^{\arityof{\ppred}} x_j = \atpos{x}{e_i}_j *
  \atpos{\psi}{e_i}_i * \Bigstar_{\ell=1}^{k_i}
  \charform{\tree_{i,\ell}}[x_1/\atpos{z}{e_i}_{\ell,1}, \ldots,
  x_{\arityof{\ppred_{i,\ell}}}/\atpos{z}{e_i}_{\ell,\arityof{\ppred_{i,\ell}}}]
  \label{eq:charform}
\end{align}

\vspace*{-\baselineskip}\noindent where each edge $e_i$,
$i\in\interv{1}{m}$, attached to the root $r$ of $\tree$ (i.e.,
$\edgerelof{\tree}(e_i)_1 = r$) is labeled by a productive rule of the
form \(\qpred_i(x_1,\ldots,x_{\arityof{\ppred}}) \leftarrow \exists
y_1 \ldots \exists y_{n_i} ~.~ \psi_i * \Bigstar_{j=1}^k
\ppred_{i,j}(z_{j,1},\ldots,z_{j,\arityof{\ppred_j}})\), such that
$\psi_i$ is a qpf formula and $\tree_{i,\ell}$ is the subtree of
$\tree$ rooted in $\edgerelof{\tree}(e_i)_{\ell+1}$, for all $\ell \in
\interv{1}{k_i}$. Note that $\charform{\tree}$ is a qpf formula, for
each $\tree\in\parsetreeof{\asid}{\ppred}$.

\ifLongVersion
\else
We recall our purpose, which is finding a fragment of \slr, for which
the sets $\sidsem{\apred}{\asid}$ have bounded tree-width and are
\mso-definable. First, we identify a set of \emph{canonical}
$\asid$-models of $\apred$, in which the existentially quantified
variable are instantiated with distinct vertices, and prove that these
models form a \hr\ set (\S\ref{sec:canonical}). Moreover, if $\asid$
is regular, the graph grammar that defines this set is regular (Lemma
\ref{lemma:canonical-hr}), thus producing \mso-definable sets.
Second, any set $\sidsem{\apred}{\asid}$ results from a \emph{fusion}
operation applied to the set of canonical $\asid$-models
(Def. \ref{def:fusion}). Intuitively, the fusion joins several
vertices of a simple c-graph into one, provided that the result
remains a simple c-graph. The inverse of fusion, called \emph{fission}
splits the vertices of a simple c-graph into sets of distinct copies
and redirects each edge incident to a split vertex to one of its
copies.

The main idea is to express fission as an \emph{\mso-definable
  transduction}, that creates first a constant number of disjoint
copies of the input graph and extracts, from all the copies at once, a
graph that can be fusioned back into the input graph. The argument for
the \mso-definability follows from the Backwards Translation Theorem
(Thm. \ref{thm:bt}), which states that the inverse image of an
\mso-definable set via an \mso-definable transduction is
\mso-definable. Since $\sidsem{\apred}{\asid}$ is the inverse image of
the \mso-definable set of canonical $\asid$-models of $\apred$ via
fission, we establish that $\sidsem{\apred}{\asid}$ is \mso-definable.

Unfortunately, it is currently an open problem, whether fission is
\mso-definable, in general\footnote{We show that fission is not
  \msone-definable, for the fragment \msone\ of \mso\ with
  quantification restricted to sets of vertices only, see Appendix
  \ref{app:fission}.}. We define fission in \mso, under the
restriction that $\sidsem{\apred}{\asid}$ can be generated from the
canonical models, by applying a bounded number of elementary fusions,
that join two vertices (\S\ref{sec:fission}). This \emph{bounded
  fusionability} condition also ensures that $\sidsem{\apred}{\asid}$
has bounded tree-width (Lemma \ref{lemma:bf-btw}).

Finally, we provide easy-to-check syntactic conditions, called
\emph{rigidity}, that guarantees bounded fusionability of the set of
canonical $\asid$-models of $\apred$ (\S\ref{sec:rigidity}). In
addition to regularity, the rigidity condition defines an expressive
fragment of \slr\ that combines bounded tree-width with
\mso-definabilty and in which entailments between formul{\ae} are
decidable. We state below our main result:
\fi

\begin{theoremE}[][category=proofs]\label{thm:main}
  Let $\asid$ be a regular and rigid SID for a nullary predicate
  $\apred$. Then, $\sidsem{\apred}{\asid}$ has bounded tree-width, for
  an effectively computable bound, and is \mso-definable.
\end{theoremE}
\begin{proofE}
  Without losing generality, we assume that $\asid$ is equality-free
  and all-satisfiable, by Lemmas \ref{lemma:eq-free} and
  \ref{lemma:all-sat}. Since $\asid$ is rigid for $\apred$, we have
  $\fbof{\rcsem{\apred}{\asid}} \leq B$, for an effectively computable
  bound $B \geq 0$, by Lemma \ref{lemma:rigid-bf}.  Hence
  $\sidsem{\apred}{\asid} = \fusion{B}(\rcsem{\apred}{\asid})$, by
  Lemma \ref{lemma:fusion}. Since $\csem{\apred}{\asid}$ is generated
  by a \hr\ grammar $\grammar$, by Lemma \ref{lemma:canonical-hr}, a
  local change of the substitution operations in $\grammar$ generates
  $\rcsem{\apred}{\asid}$ (we add $\diseq$-labeled edges to all
  c-graphs $\graph$ in the rules $\gamma(\arule)$, such that there is
  a disequality in $\arule$). Then, $\twof{\rcsem{\apred}{\asid}}\leq
  K$, where $K$ is the maximum number of sources in $\grammar$, by a
  classical result, see, e.g.,~\cite[Proposition
    4.7]{courcelle_engelfriet_2012}. By Lemma \ref{lemma:bf-btw}, we
  obtain $\twof{\apred}{\asid} \leq K + B$. Because $\asid$ is
  regular, we obtain that $\rcsem{\apred}{\asid}$ is \mso-definable,
  by Lemma \ref{lemma:rich-canonical-mso}. Since $\fission{B}$ is an
  \mso-definable transduction, by Lemma \ref{lemma:fission} and
  Prop. \ref{prop:comp-trans}, we conclude that
  $\sidsem{\apred}{\asid} = (\fission{B})^{-1}(\rcsem{\apred}{\asid})$
  is \mso-definable, by Theorem \ref{thm:bt}. \qed
\end{proofE}

\section{Canonical Models}
\label{sec:canonical}

We fix a regular SID $\asid$ and a a nullary predicate $\apred$, in
the following. A model is \emph{canonical} iff it can be defined using
a store that matches only those variables that are explicitly equated
in the unfolding of the rules that produced the model. For a qpf
formula $\phi$, we write $x \eqof{\phi} y$ (resp. $x \not\eqof{\phi}
y$) iff $x=y$ is (resp. is not) a logical consequence of $\phi$. A
store $\store$ is \emph{canonical for $\phi$} iff
$\store(x)=\store(y)$ only if $x \eqof{\phi} y$, for all
$x,y\in\fv{\phi}$.

Let $\alphabet$ be a fixed alphabet of edge labels to which we add a
distinguished binary edge label $\diseq$, that records disequalities,
and denote by $\alphabet^\diseq$ the set $\alphabet \cup
\set{\diseq}$. For a c-graph $\graph$, we denote by
$\proj{\graph}{\alphabet}$ the c-graph obtained by deleting all edges
with label $\diseq$ from $\graph$.

\begin{definition}\label{def:canonical-model}
  A \emph{rich canonical $\asid$-model} of $\apred$ is a c-graph
  $\graph \in \cgraphsof{\alphabet^\diseq}$, for which there exists a
  parse tree $\tree\in\parsetreeof{\asid}{\apred}$ and a store
  $\store$, canonical for $\charform{\tree}$, such
  that: \begin{compactenum}
  \item\label{it1:def:canonical-model} $\proj{\graph}{\alphabet} \models^\store
    \charform{\tree}$ and,
  \item\label{it2:def:canonical-model} for all
    $u,v\in\vertof{\graph}$, there exists an edge $e \in
    \edgeof{\graph}$ such that $\labof{\graph}(e)=\diseq$ and
    $\edgerelof{\graph}(e) = \tuple{u,v}$ iff there exist variables $x
    \in \store^{-1}(u)$ and $y \in \store^{-1}(v)$ such that $x \neq
    y$ occurs in $\charform{\tree}$.
  \end{compactenum}
  We denote by $\rcsem{\apred}{\asid}$ the set of rich canonical
  $\asid$-models of $\apred$ and
  $\csem{\apred}{\asid}\isdef\set{\proj{\graph}{\alphabet} \mid
    \graph\in\rcsem{\apred}{\asid}}$ the set of \emph{canonical}
  $\asid$-models of $\apred$.
\end{definition}

\begin{textAtEnd}[category=proofs]
  We show below that the set of canonical $\asid$-models is closed
  under isomorphism:
\begin{lemma}\label{lemma:canonical-isomorphism}
  Let $\graph,H\in \cgraphsof{\alphabet}$ be isomorphic c-graphs.
  Then, $\graph \in \csem{\apred}{\asid} \iff H \in
  \csem{\apred}{\asid}$.
\end{lemma}
\begin{proof}
  Assume that $\graph \in \csem{\apred}{\asid}$ (the other
  direction is symmetric). Then, there exists a parse tree
  $\tree\in\parsetreeof{\asid}{\apred}$ and a store $\store$,
  canonical for $\charform{\tree}$, such that
  $\graph\models^\store\charform{\tree}$. Let $h : \vertof{\graph}
  \cup \edgeof{\graph} \rightarrow \vertof{H} \cup \edgeof{H}$ be the
  bijection that establishes the isomorphism between $\graph$ and
  $H$. Then, $\store \circ h$ is canonical for $\charform{\tree}$ and
  $H \models^{\store \circ h} \charform{\tree}$ is proved by induction
  on the structure of $\charform{\tree}$. Actually, the more general
  statement holds $\graph \models^\store \phi \Rightarrow H
  \models^{\store\circ h} \phi$, for any \slr\ formula $\phi$. Hence,
  we obtain $H \in \csem{\apred}{\asid}$.
\end{proof}
\end{textAtEnd}

Rich canonical models are important, because each $\asid$-model of
$\apred$ can be obtained by applying a \emph{fusion} to a rich
canonical model of $\apred$, i.e., a unary operation that glues
together vertices. This operation is defined using quotienting:

\begin{definition}\label{def:quotient}
  Let $\graph = \tuple{\vertof{\graph}, \edgeof{\graph},
    \labof{\graph}, \edgerelof{\graph}} \in
  \cgraphsof{\alphabet^\diseq}$ be a simple c-graph of type $0$. An
  equivalence relation $\approx \subseteq \vertof{\graph} \times
  \vertof{\graph}$ is \emph{compatible} with $\graph$ iff the
  following hold: \begin{compactenum}
  \item\label{it1:def:quotient} $\edgerelof{\graph}(e_1) \not\approx
    \edgerelof{\graph}(e_2)$, for all edges
    $e_1 \neq e_2\in\edgeof{\graph}$, such that
    $\labof{\graph}(e_1)=\labof{\graph}(e_2)$,
    %
    %
  \item\label{it2:def:quotient} for each edge $e\in\edgeof{\graph}$,
    such that $\labof{\graph}(e)=\diseq$ and
    $\edgerelof{\graph}(e)=\tuple{v_1,v_2}$, we have $v_1 \not\approx
    v_2$.
  \end{compactenum}
  The \emph{quotient} of $\graph$ is the c-graph $\graph_{/\approx}$,
  where $\vertof{\graph_{/\approx}} \isdef \set{[v]_{/\approx} \mid v
    \in \vertof{\graph}}$, $\edgeof{\graph_{/\approx}} \isdef
  \edgeof{\graph}$, $\labof{\graph_{/\approx}}(e) \isdef
  \labof{\graph}(e)$, $\edgerelof{\graph_{/\approx}}(e) \isdef
  \tuple{[v_1]_{/\approx}, \ldots, [v_k]_{/\approx}}$, if
  $\edgerelof{\graph}(e) = \tuple{v_1,\ldots,v_k}$, for all $e \in
  \edgeof{\graph_{\approx}}$, and $\sourceof{\graph_{/\approx}}(i)
  \isdef [\sourceof{\graph}(i)]_{/\approx}$, for all $i \in
  \interv{1}{n}$, where $[v]_{/\approx}$ denotes the
  $\approx$-equivalence class of $v\in\vertof{\graph}$.
\end{definition}
The quotients of simple graphs w.r.t. compatible relations are simple
c-graphs as well. An equivalence relation $\approx$ is
\emph{generated} by a set of pairs $\set{(u_1,v_1), \ldots,
  (u_k,v_k)}$ iff it is the least equivalence relation such that $u_i
\approx v_i$, for all $i \in \interv{1}{k}$.  We say that $\approx$ is
$k$-generated if $k$ is the minimal cardinality of a set of pairs that
generates $\approx$.

\begin{definition}\label{def:fusion}
  The \emph{fusion} of a simple c-graph
  $\graph\in\cgraphsof{\alphabet^\diseq}$ is the set of simple
  c-graphs $\transfusion(\graph)$ isomorphic to $\graph_{/\approx}$,
  where $\approx$ is any equivalence relation compatible with
  $\graph$. Conversely, the \emph{fission} operation is defined as
  $\transfission(\graph) \isdef \set{\graph' \mid \graph \in
    \transfusion(\graph')}$. We write $\fusion{k}$ and $\fission{k}$
  when each equivalence relation $\approx$ in $\transfusion$ is
  $k$-generated. These operations are lifted from c-graphs to graphs
  in the usual way.
\end{definition}
The following lemma shows that each $\asid$-model of $\apred$ is the
result of applying fusion to a rich canonical $\asid$-model thereof:

\begin{lemma}[Lemma 7 in \cite{DBLP:journals/corr/abs-2310-09542}]\label{lemma:fusion}
  $\sidsem{\apred}{\asid} = \proj{\transfusion(\rcsem{\apred}{\asid})}{\alphabet}$.
\end{lemma}

For a c-graph $\graph$ we define its \emph{fusion bound}
$\fbof{\graph}$ as the maximum integer $k$ for which there exists an
equivalence $\approx$ compatible with $\graph$ and $k$-generated. We
say that a set of c-graphs $\graphs$ is \emph{bounded fusionable}
iff the set $\set{ \fbof{\graph} \mid \graph \in \graphs}$ is finite.

\begin{lemmaE}[][category=proofs]\label{lemma:bf-btw}
  $\twof{\sidsem{\apred}{\asid}} \leq \twof{\rcsem{\apred}{\asid}} + \fbof{\rcsem{\apred}{\asid}}$.
\end{lemmaE}
\begin{proofE}
  We will show that, for a c-graph $\graph$ and an equivalence
  relation $\approx \subseteq \vertof{\graph} \times \vertof{\graph}$
  compatible with $\graph$ which is $k$-generated it holds
  $\twof{\graph_{/\approx}} \le \twof{\graph} + k$.

  For a $k$-generated equivalence there are at most $k$ equivalence
  classes with 2 or more elements.  Let $\ell \le k$ be the number of such
  classes and let $u_i \in \vertof{G}$ be some chosen representative
  in each class, for $i\in\interv{1}{\ell}$.

  Let consider an arbitrary tree decomposition $(\tree,\beta)$ for
  $\graph$.  Then, construct the tree decomposition $(\tree', \beta')$
  for $\graph$ by taking $\tree' \isdef \tree$ and $\beta'(n) \isdef
  \beta(n) \cup \set{u_i~|~i \in\interv{1}{\ell}}$ for every node $n$.
  Obviously, $\width{\tree'} \le \width{\tree} + k$.  Now, construct the tree
  decomposition $(\tree'', \beta'')$ for $\graph_{/\approx}$ by taking
  $\tree'' \isdef \tree'$ and $\beta''(n) \isdef \set{ [u]_{/\approx}
    \mid u \in \beta(n)}$ for every node $n$.  It is an easy check
  that $(\tree'',\beta'')$ is actually a valid tree decomposition for
  $\graph_{/\approx}$, in particular, the connectivity property is
  guaranteed because the representatives of the non-trivial
  equivalence classes were propagated everywhere in $\tree''$.  Also,
  by construction we have $\width{\tree''} \le \width{\tree'}$.

  Altogether, we proved that for every tree decomposition
  $(\tree,\beta)$ for $\graph$ there exists a tree decomposition
  $(\tree'',\beta'')$ of $\graph_{/\approx}$ such that $\width{\tree''}
  \le \width{\tree} + k$.  That implies $\twof{\graph_{/\approx}} \le
  \twof{\graph} + k$. \qed
\end{proofE}

\noindent
A consequence is that $\sidsem{\apred}{\asid}$ has bounded tree-width
if $\rcsem{\apred}{\asid}$ is bounded fusionable. This is because
$\rcsem{\apred}{\asid}$ is a \hr\ set (a consequence of Lemma
\ref{lemma:canonical-hr}) and \hr\ sets are known to have bounded
tree-width, by a classical result, e.g.,~\cite[Proposition
  4.7]{courcelle_engelfriet_2012}. The dual statement is not true, a
counterexample being the SID from Fig. \ref{fig:sids} (b), where
$\twof{\sidsem{\apred}{\asid}} \leq 2$, yet $\sidsem{\apred}{\asid}$
is not bounded fusionable, i.e., the vertices on the second level can
be joined among themselves and with the ones from the first level.

We simplify the technical development by assuming that no equalities
occur in $\asid$ and show that this loses no generality (Lemma
\ref{lemma:eq-free}).

\begin{definition}\label{def:eq-free}
  A formula is \emph{equality-free} iff it contains no equalities nor
  predicate atoms in which the same variable occurs twice. A rule
  $\apred(x_1,\ldots,x_n) \leftarrow \phi$ is equality-free iff $\phi$
  is equality-free. A SID is equality-free iff it consists of
  equality-free rules.
\end{definition}

\begin{lemma}[Lemma 9 in \cite{Concur23}]\label{lemma:eq-free}
  Given a SID $\asid$, one can build an equality-free SID
  $\overline{\asid}$, such that $\sidsem{\apred}{\asid} =
  \sidsem{\apred}{\overline{\asid}}$, for each nullary predicate
  $\apred$.
\end{lemma}

Moreover, we assume that each $(\asid,\apred)$-parse tree has a
satisfiable characteristic formula. Again, this assumption loses no
generality (Lemma \ref{lemma:all-sat}).

\begin{definition}\label{def:all-sat}
  A SID $\asid$ is \emph{all-satisfiable for $\apred$} iff
  $\charform{\tree}$ is a satisfiable predicate-free formula, for each
  parse tree $\tree\in\parsetreeof{\asid}{\apred}$.
\end{definition}

\begin{lemma}[Lemma 17 in \cite{DBLP:journals/corr/abs-2310-09542}]\label{lemma:all-sat}
  For each SID $\asid$ and nullary predicate $\apred$, one can build a
  SID $\overline{\asid}$ all-satisfiable for $\apred$, such that: \begin{compactenum}
  \item\label{it1:lemma:all-sat} for each parse tree $\tree\in\parsetreeof{\asid}{\apred}$,
    such that $\charform{\tree}$ is satisfiable, there exists
    a parse tree $\overline{\tree}\in\parsetreeof{\overline{\asid}}{\apred}$, such
    that $\charform{\tree}$ and $\charform{\overline{\tree}}$ are equivalent, and
  \item\label{it2:lemma:all-sat} for each parse tree
    $\overline{\tree}\in\parsetreeof{\overline{\asid}}{\apred}$, there
    exists a parse tree $\tree\in\parsetreeof{\asid}{\apred}$, such that
    $\charform{\tree}$ and $\charform{\overline{\tree}}$ are equivalent.
  \end{compactenum}
\end{lemma}

We show another important property concerning canonical models, that
is obtained following the intuition of the correspondence between SIDs
and \hr\ grammars (Def. \ref{def:regular-sid}). 

\begin{lemmaE}[][category=proofs]\label{lemma:canonical-hr}
  Let $\asid$ be an all-satisfiable, equality-free regular SID. Then
  one can build a regular graph grammar $\grammar=(\nonterm,\rules)$,
  with a distinguished nonterminal $w\in\nonterm$ and a bijection
  $\gamma:\asid\rightarrow\rules$ such that, for each tree $\tree$,
  $\tree \in \parsetreeof{\asid}{\apred} \iff
  \gamma(\tree)\in\parsetreeof{\grammar}{w}$, where $\gamma(\tree)$ is
  the tree obtained by changing the edge label $\labof{\tree}(e)$ into
  $\gamma(\labof{\tree}(e))$, for all $e \in
  \edgeof{\tree}$. Consequently, we have $\csem{\apred}{\asid} =
  \langof{w}{\grammar}$.
\end{lemmaE}
\begin{proofE}
  The graph grammar $\grammar=(\nonterm,\rules)$ and the
  bijective mapping $\gamma:\asid\rightarrow\rules$ are defined as
  follows: \begin{compactitem}
  \item $\nonterm$ is the set of predicates that occur in $\asid$,
  \item for each rule $\arule$ of $\asid$ of the form:
    \begin{align*}
      \bpred_0(x_1, \ldots, x_{\arityof{\bpred_0}}) \leftarrow \exists
      y_1 \ldots \exists y_m ~.~ \psi * \Bigstar_{\ell=1}^k
      \bpred_\ell(z_{\ell,1}, \ldots, z_{\ell,\arityof{\bpred_\ell}})
    \end{align*}
    where $\psi$ is a qpf formula and $z_{\ell,j} \in \set{x_1,
      \ldots, x_{\arityof{\bpred_0}}} \cup \set{y_1,\ldots,y_m}$, for
    all $\ell\in\interv{1}{k}$ and $j \in
    \interv{1}{\arityof{\bpred_\ell}}$, the grammar rule
    $\gamma(\arule)$ is defined as follows: \begin{compactitem}
    \item if $m=0$, $\psi = \emp$,
      $\arityof{\bpred_\ell}=\arityof{\bpred_0} = n$ and $z_{\ell,j} =
      x_j$, for all $\ell\in\interv{1}{k}$ and $j \in \interv{1}{n}$,
      then $\gamma(\arule)$ is the rule:
      \begin{align*}
        \bpred_0 \rightarrow \bpred_1 \parallel_n \ldots \parallel_n \bpred_k
      \end{align*}
    \item otherwise, $\gamma(\arule)$ is the rule:
      \begin{align*}
        \bpred_0 \rightarrow (\graph,\tuple{\bpred_1,z_{1,1},\ldots,z_{1,\arityof{\bpred_1}}},
        \ldots, \tuple{\bpred_k,z_{k,1},\ldots,z_{k,\arityof{\bpred_k}}})
      \end{align*}
      where $\graph$ is the c-graph of type $\arityof{\bpred_0}$ defined as follows:
            \begin{align*}
      \vertof{\graph} \isdef & \set{x_1,\ldots,x_{\arityof{\bpred_0}}}\cup\set{y_1,\ldots,y_m} \\
      \edgeof{\graph} \isdef & \set{\tuple{\arel,\zeta_1,\ldots,\zeta_{\arityof{\arel}}}
        \mid \arel(\zeta_1,\ldots,\zeta_{\arityof{\arel}}) \text{ occurs
          in } \psi} ~\cup \\
      & \set{\tuple{\bpred_\ell,z_{\ell,1}, \ldots, z_{\ell,\arityof{\bpred_\ell}}} \mid \ell \in
        \interv{1}{k},~ j \in \interv{1}{\arityof{\bpred_\ell}}} \\
      \labof{\graph}(e) \isdef & \left\{\begin{array}{ll}
      \arel & \text{, if } e = \tuple{\arel,\zeta_1,\ldots,\zeta_{\arityof{\arel}}} \\
      \bpred_\ell & \text{, if } e = \tuple{\bpred_\ell,z_{\ell,1}, \ldots, z_{\ell,\arityof{\bpred_\ell}}}
      \end{array}\right. \text{, for all } e \in \edgeof{\graph} \\
      \edgerelof{\graph}(e) \isdef & \left\{\begin{array}{ll}
      \tuple{\zeta_1,\ldots,\zeta_{\arityof{\arel}}} & \text{, if }
      e = \tuple{\arel,\zeta_1,\ldots,\zeta_{\arityof{\arel}}} \\
      \tuple{z_{\ell,1}, \ldots, z_{\ell,\arityof{\bpred_\ell}}} & \text{, if }
      e = \tuple{\bpred_\ell,z_{\ell,1}, \ldots, z_{\ell,\arityof{\bpred_\ell}}}
      \end{array}\right. \text{, for all } e \in \edgeof{\graph} \\
      \sourceof{\graph}(i) \isdef & x_i \text{, for all } i \in \interv{1}{\arityof{\bpred_0}}
      \end{align*}
    \end{compactitem}
  \end{compactitem}
  We show that $\grammar$ is regular, provided that $\asid$ is
  regular, where $\mathbb{Q}$ is the set of productive predicates. In
  particular, each productive rule $\bpred_0 \rightarrow (\graph,
  \tuple{\bpred_1,z_{1,1},\ldots,z_{1,\arityof{\bpred_1}}}, \ldots,
  \langle\bpred_k,z_{k,1},\\\ldots,z_{k,\arityof{\bpred_k}}\rangle)$
  is such that $\bpred_0 \in \mathbb{Q}$ and each unproductive rule is
  either of the form $\bpred_0 \rightarrow \bpred_0 \parallel_n
  \bpred_1$ or $\bpred_0 \rightarrow \bpred_1 \parallel_n \ldots
  \parallel_n \bpred_k$, where $\bpred_0 \in \nonterm \setminus
  \mathbb{Q}$ and $\bpred_1, \ldots, \bpred_k \in
  \mathbb{Q}$. Moreover, the graph operation $(\graph,
  \tuple{\bpred_1,z_{1,1},\ldots,z_{1,\arityof{\bpred_1}}}, \ldots,
  \tuple{\bpred_k,z_{k,1},\ldots,z_{k,\arityof{\bpred_k}}})$ of each
  productive rule is regular, because: \begin{compactitem}
  \item condition (\ref{it11:regular-graph-operation}) of
    Def. \ref{def:regular-graph-grammar} follows from point
    (\ref{it1:def:regular-sid}) of Def. \ref{def:regular-sid},
  \item condition (\ref{it12:regular-graph-operation}) of
    Def. \ref{def:regular-graph-grammar} follows from point
    (\ref{it21:def:regular-sid}) of Def. \ref{def:regular-sid},
  \item condition (\ref{it2:regular-graph-operation}) of
    Def. \ref{def:regular-graph-grammar} follows from point
    (\ref{it22:def:regular-sid}) of Def. \ref{def:regular-sid}.
  \end{compactitem}
  Let the nonterminal $w$ be $\apred$. We prove that
  $\parsetreeof{\grammar}{w} = \gamma(\parsetreeof{\asid}{\apred})$,
  by proving the more general statement
  $\gamma(\parsetreeof{\asid}{\ppred}) =
  \parsetreeof{\grammar}{\ppred}$, for any $\ppred \in \preds$:

  \noindent``$\subseteq$'' Let $\tree \in \parsetreeof{\asid}{\ppred}$
  be a parse tree. We build a derivation tree $U$ of $\grammar$
  (Def. \ref{def:grammar-parse-trees}), such that $U \reduce^*
  \gamma(\tree)$, by induction on the structure of $\tree$. Let
  $e_1,\ldots,e_m$ be the edges attached to $\rootof{\tree}$. We
  distinguish the following two cases: \begin{compactitem}
  \item if $\ppred\in\mathbb{Q}$ then $m=1$ and $\labof{\tree}(e_1)$
    is a productive rule. Let $\tree_i \in
    \parsetreeof{\asid}{\ppred_i}$, for $i \in \interv{1}{k}$, be the
    subtrees of $\tree$ rooted in the vertices attached to $e_1$ at
    positions $2,\ldots,k+1$, respectively. By the inductive
    hypothesis, for each $i \in \interv{1}{k}$, there exists a
    derivation tree $U_i$, such that $U_i \reduce^* \gamma(\tree_i)
    \in \parsetreeof{\grammar}{\ppred_i}$. We define $U$ by attaching
    each $U_i$ to an edge labeled by $\gamma(\labof{\tree}(e_1))$ on
    position $i+1$, for each $i \in \interv{1}{k}$. The root of $U$ is
    a new vertex attached to this edge on the first position.
  \item else, $\ppred\in\preds\setminus\mathbb{Q}$ and there exists
    $\ell \in \interv{0}{m}$, such that: \begin{compactitem}
  \item if $\ell\geq1$ there exists a rule
    $\ppred(x_1,\ldots,x_{\arityof{\ppred}}) \leftarrow
    \Bigstar_{i=1}^\ell \qpred_i(x_1,\ldots,x_{\arityof{\qpred_i}})$,
    with $\ppred\in\mathbb{Q}$, such that $\labof{\tree}(e_i)$ is a
    rule with left-hand side
    $\qpred_i(x_1,\ldots,x_{\arityof{\qpred_i}})$, for all $i \in
    \interv{1}{\ell}$. Let $\tree_{i}$ be the subtree of $\tree$
    rooted in $\rootof{\tree}$ with $e_i$ as the only attached edge,
    for all $i \in \interv{1}{\ell}$. By the inductive hypothesis,
    there exist derivation trees $U_{i}$ of $\grammar$, such that
    $U_{i} \reduce^* \gamma(T_{i}) \in
    \parsetreeof{\grammar}{\qpred_i}$, for all $i \in
    \interv{1}{\ell}$. Moreover, by definition, $\grammar$ has a rule
    $\ppred \rightarrow \qpred_1 \parallel_n \ldots \parallel_n
    \qpred_\ell$. Let $V_\ell$ be the derivation tree of $\grammar$
    obtained by attaching $U_1, \ldots, U_\ell$ to the positions
    $2,\ldots,\ell+1$ of a new edge labeled with this rule, where
    $\rootof{V_\ell}$ is a fresh vertex attached to the edge on its
    first position.
  \item for each $j \in \interv{\ell+1}{m}$, there exists a rule
    $\ppred(x_1,\ldots,x_{\arityof{\ppred}}) \leftarrow
    \ppred(x_1,\ldots,x_{\arityof{\ppred}}) *
    \qpred(x_1,\ldots,x_{\arityof{\qpred}})$, with
    $\ppred\in\mathbb{Q}$, such that the rule $\labof{\tree}(e_j)$ has
    $\qpred(x_1,\ldots,x_{\arityof{\qpred}})$ as left-hand side. Let
    $\tree_j$ be the subtree of $\tree$ rooted in $\rootof{\tree}$,
    with $e_j$ as the only attached edge, for all $j \in
    \interv{\ell+1}{m}$. By the inductive hypothesis, there exists a
    derivation tree $U_j$ of $\grammar$, such that $U_j \reduce^*
    \gamma(T_j) \in \parsetreeof{\grammar}{\qpred}$, for all $j \in
    \interv{\ell+1}{m}$. Moreover, by definition, $\grammar$ has a
    rule $\ppred \rightarrow \ppred \parallel_n \qpred$. For each $j
    \in \interv{\ell+1}{m}$, we define each tree $V_j$ by attaching to
    an edge labeled by this rule the tree $V_{j-1}$ on position $2$,
    $U_j$ on position $3$ and a fresh root on its first position.
  \end{compactitem}
    Let $U$ be the tree $V_m$. It is easy to show that $U \reduce^*
    \gamma(\tree)$, obtained by successively removing the edges
    labeled with unproductive rules from $V_m$.
  \end{compactitem}

  \noindent``$\supseteq$'' Let $U \in \parsetreeof{\grammar}{\ppred}$
  be a parse tree. We build a derivation tree $V$ of $\grammar$ such
  that $V \reduce^* U$. The construction of the parse tree $\tree \in
  \parsetreeof{\asid}{\ppred}$ such that $U = \gamma(\tree)$ is by
  induction on the structure of $V$. Let $f_1,\ldots,f_m$ be the edges
  attached to $\rootof{U}$, $\qpred_i\in\mathbb{Q}$ be the left-hand
  side of $\labof{U}(f_i)$ and $U_i$ be the subtrees of $U$ rooted in
  $\rootof{U}$, with $f_i$ as the only edge attached, for $i \in
  \interv{1}{m}$. By the definition of $\grammar$, there exists $\ell
  \in \interv{0}{m}$ and rules $p_\ell : \ppred \leftarrow \qpred_1
  \parallel_n \ldots \parallel_n \qpred_\ell$ and $p_j : \ppred
  \leftarrow \ppred \parallel_n \qpred_j$, for all $j \in
  \interv{\ell+1}{m}$, with $\ppred\in\nonterm\setminus\mathbb{Q}$ (if
  necessary, we reorder the nonterminals $\qpred_1, \ldots,
  \qpred_m$). We define $V_\ell$ by attaching $U_1, \ldots, U_\ell$ on
  positions $2, \ldots, \ell+1$ to an edge labeled by $\ppred
  \leftarrow \qpred_1 \parallel_n \ldots \parallel_n \qpred_\ell$ and
  a fresh root on position $1$. For each $j \in \interv{\ell+1}{m}$,
  we define successively $V_j$ by attaching $V_{j-1}$ and $U_j$ to an
  edge labeled by $\ppred \leftarrow \ppred \parallel_n \qpred_j$ on
  positions $2$ and $3$ and a fresh root on position $1$. Let $V =
  V_m$. It is easy to prove that $V \reduce^* U$. By the inductive
  hypothesis, there exists $\tree_i \in
  \parsetreeof{\asid}{\qpred_i}$, such that $U_i = \gamma(\tree_i)$,
  for each $i \in \interv{1}{m}$. We define $\tree$ by joining the
  roots of $\tree_1,\ldots,\tree_m$ and prove that $\tree \in
  \parsetreeof{\asid}{\ppred}$ by the existence of unproductive rules
  $\gamma^{-1}(p_j) \in \asid$, for all $j \in \interv{\ell}{m}$,
  following from the definitions of $\grammar$ and $\gamma$.

  \vspace*{.5\baselineskip}\noindent We are left with proving that
  $\csem{\apred}{\asid}=\langof{w}{\grammar}$. ``$\subseteq$'' Let
  $\graph \in \csem{\apred}{\asid}$ be a c-graph. By
  Def. \ref{def:canonical-model}, there exists a parse tree $\tree \in
  \parsetreeof{\asid}{\apred}$ and a store $\store$, canonical for
  $\charform{\tree}$, such that $\graph \models^\store
  \charform{\tree}$. Since $\asid$ is equality-free, the mapping
  $\proj{\store}{\fv{\charform{\tree}}}$ is a bijection. By the
  previous point, there exists a parse tree $U \in
  \parsetreeof{\grammar}{w}$, such that $\tree=\gamma(U)$. Using the
  definition of $\grammar$ and $\gamma$, one can show that
  $\proj{\store}{\fv{\charform{\tree}}}$ establishes an isomorphism
  between $\eval{U}$ and $\graph$, hence
  $\graph\in\langof{w}{\grammar}$. ``$\supseteq$'' Let
  $\graph\in\langof{w}{\grammar}$ be a c-graph. Then, there exists a
  parse tree $U \in \parsetreeof{\grammar}{w}$ such that
  $\graph=\eval{U}$. By the previous point, there exists a parse tree
  $\tree\in\parsetreeof{\asid}{\apred}$, such that
  $U=\gamma(\tree)$. Since $\asid$ is all-satisfiable
  (Def. \ref{def:all-sat}), there exists a store $\store$ and a graph
  $H$, such that $H \models^\store \charform{\tree}$.  Moreover,
  because $\asid$ is equality-free, we can assume w.l.o.g. that
  $\proj{\store}{\fv{\charform{\tree}}}$ is bijective, thus canonical
  for $\charform{\tree}$, i.e., $H \in \csem{\apred}{\asid}$. By the
  definition of $\grammar$ and $\gamma$, one shows that
  $\proj{\store}{\fv{\charform{\tree}}}$ establishes an isomorphism
  between $\graph$ and $H$. Hence, $\graph \in \csem{\apred}{\asid}$,
  because $\csem{\apred}{\asid}$ is closed under isomorphism, by Lemma
  \ref{lemma:canonical-isomorphism}. \qed
\end{proofE}
As an immediate consequence, we obtain that $\csem{\apred}{\asid}$ is
\mso-definable, because each language produced by a regular
\hr\ grammar is \mso-definable. We extend this result to rich
canonical models:

\begin{lemmaE}[][category=richcanon]\label{lemma:rich-canonical-mso}
  $\rcsem{\apred}{\asid}$ is \mso-definable.
\end{lemmaE}
\begin{proofE}
  Consider the \mso-definable
  $(\alphabet^\diseq,\alphabet)$-transduction $D$ that removes the
  edges labeled by $\diseq$ from the input graph. By Lemma
  \ref{lemma:canonical-mso}, $\csem{\apred}{\asid}$ is
  \mso-definable. By Theorem \ref{thm:bt}, the set
  $D^{-1}(\csem{\apred}{\asid})$ is \mso-definable. We conclude the
  proof by showing that:
  \[\rcsem{\apred}{\asid} = D^{-1}(\csem{\apred}{\asid}) ~\cap~ \set{\graph \mid \graph \Models
    \forall x \forall y ~.~ \diseqform(x,y) \leftrightarrow \exists z
    ~.~ \edgrel{\diseq}(z,x,y)}\]

  \noindent''$\subseteq$'' Let $\graph \in \rcsem{\apred}{\asid}$ be a
  c-graph. Then $\proj{\graph}{\alphabet} \in \csem{\apred}{\asid}$
  and $\graph \in D^{-1}(\csem{\apred}{\asid})$ follows. Hence
  $\proj{\graph}{\alphabet} \Models^{\overline\store}
  \charform{\tree}$, for some parse tree $\tree =
  \defdof{\scheme^\trparse_\asid}{\store}(\graph)$, where $\store$ is
  a store that interprets the parameters $\mathcal{W}$ and
  $\overline\store$ is a canonical store for $\charform{\tree}$. Let
  $v_1,v_2 \in \vertof{\graph}$ be arbitrary vertices. Then, $\graph
  \Models^{[x\leftarrow v_1, y\leftarrow v_2]} \exists z_0 ~.~
  \edgrel{\diseq}(z,x,y)$ iff there exists an edge $e \in
  \edgeof{\graph}$ with label $\labof{\graph}(e)=\diseq$ and endpoints
  $\edgerelof{\graph}(e)=\tuple{v_1,v_2}$. Since $\graph \in
  \rcsem{\apred}{\asid}$, this is the case iff there exists an edge $f
  \in \edgeof{\tree}$ such that the disequality $\atpos{z}{f}_1 \neq
  \atpos{z}{f}_2$ occurs within $\charform{\tree}$ and
  $\overline\store(\atpos{z}{f}_j)=v_j$, for all $j=1,2$. Suppose
  w.l.o.g. that $\atpos{z}{f}_1 \neq \atpos{z}{f}_2$ has been
  introduced by a productive rule $\arule$ of $\asid$, i.e., $z_1 \neq
  z_2$ occurs in the qpf formula $\psi$ of $\arule$. Let $\set{v_0}
  \isdef \store(C) \cap \overline\store(\fv{\atpos{\psi}{f}})$ be the
  distinguished vertex introduced by $f$. Moreover, suppose
  w.l.o.g. that $f$ is an $i$-edge of $\tree$. By Corollary
  \ref{cor:regular-sid-parsable} and Lemma
  \ref{lemma:regular-sid-internal} the latter condition is equivalent
  to:
  \[\graph \Models^{\store[z_0 \leftarrow v_0]} \varphi^\trparse(\mathcal{W}) \wedge
  \xi_{i,\arule,z_1}(\mathcal{W},z_0,x) \wedge
  \xi_{i,\arule,z_2}(\mathcal{W},z_0,y)\] Since the choices of $e$,
  $v_1$ and $v_2$ were arbitrary, we obtain $\graph \Models \forall x
  \forall y ~.~ \diseqform(x,y) \leftrightarrow \exists z ~.~
  \edgrel{\diseq}(z,x,y)$.

  \vspace*{.5\baselineskip} \noindent''$\supseteq$'' Let $\graph$ be a
  c-graph, such that $D(\graph) \in\csem{\apred}{\asid}$ and $\graph
  \Models \forall x \forall y ~.~ \diseqform(x,y) \leftrightarrow
  \exists z ~.~ \edgrel{\diseq}(z,x,y)$. Since
  $D(\graph)=\proj{\graph}{\alphabet}$, it remains to show that, for
  any $v_1,v_2 \in \vertof{\graph}$, we have $\graph
  \Models^{[x\leftarrow v_1, y\leftarrow v_2]} \diseqform(x,y)$ iff
  there exists an edge $e \in \edgeof{\graph}$ such that
  $\labof{\graph}(e)=\diseq$ and
  $\edgerelof{\graph}(e)=\tuple{v_1,v_2}$. By
  Def. \ref{def:canonical-model} (\ref{it1:def:canonical-model}),
  there exists a parse tree $\tree\in\parsetreeof{\asid}{\apred}$,
  such that $\proj{\graph}{\alphabet}\models^{\overline\store}
  \charform{\tree}$, for a store $\overline\store$ that is canonical
  for $\charform{\tree}$. By Corollary \ref{cor:regular-sid-parsable},
  $\tree=\defdof{\scheme^\trparse_\asid}{\store}(\graph)$, for a store
  that inteprets the parameters $\mathcal{W}$. By
  Def. \ref{def:canonical-model} (\ref{it2:def:canonical-model}), the
  latter condition is equivalent to the existence of an $i$-edge $f
  \in \edgeof{\tree}$ with label $\labof{\tree}(f)=\arule$ and of a
  disequality $\atpos{z}{f}_1 \neq \atpos{z}{f}_2$ introduced by
  $\arule$, such that $v_j = \overline\store(\atpos{z}{f}_j)$, for all
  $j=1,2$. Let $\psi$ be the qpf formula that occurs in $\arule$
  (\ref{rule:prod}) and $\set{v_0} \isdef \store(C) \cap
  \overline\store(\fv{\atpos{\psi}{f}})$ be the distinguished vertex
  introduced by $f$. Then, $\graph \Models^{\store[x\leftarrow v_1,y
      \leftarrow v_2, z_0 \leftarrow v_0]}
  \varphi^\trparse_\asid(\mathcal{W}) \wedge
  \xi_{i,\arule,z_1}(\mathcal{W},z_0,x) \wedge
  \xi_{i,\arule,z_2}(\mathcal{W},z_0,y)$, by Lemma
  \ref{lemma:regular-sid-internal}. We obtain, equivalently, that
  $\graph \Models^{[x\leftarrow v_1, y \leftarrow v_2]}
  \diseqform(x,y)$. Since the choice of $v_1$ and $v_2$ was arbitrary,
  we obtain $\graph \in \rcsem{\apred}{\asid}$, by
  Def. \ref{def:canonical-model}. \qed
\end{proofE}

\noindent The main difficulty in the proof of the above lemma is
giving the \mso\ definition of the disequality relation from a rich
canonical model (Def. \ref{def:canonical-model}), i.e., the set of
pairs of vertices attached to some $\diseq$-labeled edge.  This
technical point uses details from a proof by Courcelle~\cite[Theorem
  5.2]{CourcelleV}, deferred to Appendix \ref{app:transductions}, for
clarity reasons.

\section{\mso-Definable Transductions}
\label{sec:transductions}

The development of an \mso-definable fragment of \slr\ leverages from
the notion of \emph{\mso-definable transductions}. For
self-containment, we recall here the basic notions, slightly adapted
for our purposes. In particular, we define transductions between
graphs, instead of relational structures, as in the standard
literature (e.g.,~\cite[Definition 7.1]{courcelle_engelfriet_2012}).

Given alphabets of edge labels $\alphabet$ and $\betabet$, a
$(\alphabet,\betabet)$-\emph{transduction} is a relation $\trans
\subseteq \cgraphsof{\alphabet} \times \cgraphsof{\betabet}$. The
transduction first makes $k\geq1$ copies of the input c-graph, called
\emph{layers}, then defines output c-graphs within the $k$-times
disjoint union of these copies. If $k=1$, we say that $\trans$ is
\emph{copyless}. Let $\alphabet \otimes k \isdef
\set{(a,i_1,\ldots,i_{\arityof{a}+1}) \mid a\in\alphabet,~ i_1,
  \ldots, i_{\arityof{a}+1} \in \interv{1}{k}}$. The outcome of the
transduction depends on the valuation of zero or more
\emph{parameters} $X_1, \ldots, X_n$ interpreted as sets of vertices
and edges of the input c-graph. A
$(\alphabet,\betabet)$-\emph{transduction scheme} is a tuple of
\mso\ formul{\ae}
$\scheme=\tuple{\varphi,\set{\psi_i}_{i\in\interv{1}{k}},
  \set{\theta_{\overline{b}}}_{\overline{b}\in\betabet \otimes k}}$,
written using the relation symbols $\edgrel{a}$, for all $a \in
\alphabet$, where the formula $\varphi(X_1, \ldots, X_n)$ defines the
domain of the transduction, the formul{\ae} $\psi_i(x_1,X_1, \ldots,
X_n)$ define the $i$-th layer of the output and the formul{\ae}
$\theta_{(b,i_1,\ldots,i_{\arityof{b}+1})}(x_1, \ldots,
x_{\arityof{b}+1}, X_1, \ldots, X_n)$ define the edges with label $b$
in the output. For all
$(b,i,j_1\ldots,j_{\arityof{b}}),(c,i,k_1,\ldots,k_{\arityof{c}})\in\betabet\otimes
k$, we require that, for any c-graph $\graph$ and store $\store$: 
\begin{align}\label{eq:trans}
  \graph\Models^\store
  \theta_{(b,i,j_1,\ldots,j_{\arityof{b}})}(x,y_1,\ldots,y_{\arityof{b}})
  \wedge
  \theta_{(c,i,k_1,\ldots,k_{\arityof{c}})}(x,z_1,\ldots,z_{\arityof{c}})
  \\
   \Longrightarrow~ b=c,~ j_\ell =k_\ell,~ \store(y_\ell)=\store(z_\ell)
  \text{, for all } \ell\in\interv{1}{\arityof{b}} \nonumber
\end{align}

For an input c-graph $\graph \in \cgraphsof{\alphabet}$ and store
$\store$, such that $\graph \Models^\store \varphi$, the output
c-graph $\defdof{\scheme}{\store}(\graph) \in \cgraphsof{\betabet}$ is
defined as follows: \begin{compactitem}
\item $\vertof{\defdof{\scheme}{\store}(\graph)} \isdef \set{(v,i) \mid v \in \vertof{\graph},~ i \in
  \interv{1}{k},~ \graph \Models^{\store[x_1\leftarrow v]} \psi_i}$,
\item $\edgeof{\defdof{\scheme}{\store}(\graph)} \isdef \set{(e,i) \mid e \in \edgeof{\graph},~ i \in
  \interv{1}{k},~ \graph \Models^{\store[x_1\leftarrow e]} \psi_i}$,
\item for each edge $(e,i) \in \edgeof{\defdof{\scheme}{\store}(\graph)}$,
  $b \in \betabet$ and $i_1,\ldots,i_{\arityof{b}} \in
  \interv{1}{k}$:
  \begin{align*}
    \labof{\defdof{\scheme}{\store}(\graph)}(e,i)=b \text{ and }
    \edgerelof{\defdof{\scheme}{\store}(\graph)}(e,i) =
    \tuple{(v_1,i_1), \ldots, (v_{\arityof{b}},i_{\arityof{b}})}
    \\
    \iff \graph \Models^{\store[x_1 \leftarrow e, x_2\leftarrow v_1,
        \ldots, x_{\arityof{b}+1}\leftarrow v_{\arityof{b}}]}
    \theta_{(b,i,i_1,\ldots,i_{\arityof{b}})}
  \end{align*}
  Note that condition (\ref{eq:trans}) above ensures that each edge
  from the output graph has a unique label and is attached to a unique
  tuples of vertices.
\end{compactitem}
The function $\defd{\scheme} : \cgraphsof{\alphabet} \times
\pow{\cgraphsof{\betabet}}$ maps a graph $\graph$ into a set of
c-graphs $\defdof{\scheme}{\store}(\graph)$, one for each
interpretation of the parameters $X_1, \ldots, X_n$ by the store
$\store$. A transduction $\trans$ is \emph{\mso-definable} if there
exists a transduction scheme $\scheme$, such that
$\trans=\defd{\scheme}$. We lift \mso-definable transductions from
c-graphs to graphs, by taking the closure under isomorphism of
$\defd{\scheme}(\graph)$, where $\graph$ is a c-graph taken from an
input graph. The main property of \mso-definable transductions is the
Backwards Translation (BT) Theorem (e.g., \cite[Theorem
  1.40]{courcelle_engelfriet_2012}):
\begin{theorem}[Backwards Translation]\label{thm:bt}
  If $\mathcal{S} \subseteq \graphsof{\betabet}$ is an \mso-definable
  set and $\trans$ is an \mso-definable
  $(\alphabet,\betabet)$-transduction then
  $\trans^{-1}(\mathcal{S})\subseteq\graphsof{\alphabet}$ is an
  \mso-definable set.
\end{theorem}

\begin{proposition}\label{prop:comp-trans}
  The composition of \mso-definable transductions is \mso-definable.
\end{proposition}

\begin{textAtEnd}[category=transductions]
\subsection{Parsable Sets of Canonical Models}

A context-free set of graphs is \emph{parsable} if, for each graph in
the set, the corresponding parse trees of the generating graph grammar
(Def. \ref{def:grammar-parse-trees}) can be extracted from the graph
using an \mso-definable transduction. Even though not every
context-free set is parsable, an important property of regular graph
grammars is that they always generate parsable sets~\cite[Theorem
  5.2]{CourcelleV}. Here we introduce a similar notion for SIDs and
prove a similar result, leveraging from the tight connection between
regular SIDs and graph grammars, respectively (Lemma
\ref{lemma:canonical-hr}). In the rest of this section, let $\asid$ be
an all-satisfiable, equality-free and regular SID and $\apred$ be a
nullary predicate.

\begin{definition}\label{def:parsable}
A set of graphs $\graphs \subseteq \sidsem{\apred}{\asid}$ is
$\trans$-\emph{parsable} iff $\trans \subseteq \graphs \times
\parsetreeof{\asid}{\apred}$ is an \mso-definable transduction, such
that $\trans(\graph) = \set{\tree \in \parsetreeof{\asid}{\apred} \mid
  \graph \models^\store \charform{\tree} \text{, for a store }
  \store}$, for all $\graph \in \graphs$.
\end{definition}


An edge $e \in \edgeof{\graph}$ of a c-graph $\graph$ is an
\emph{$i$-edge} if there exists another edge $e' \in \edgeof{\graph}$
such that $\edgerelof{\graph}(e)_1 = \edgerelof{\graph}(e')_i$, for
some $i \in \interv{1}{\arityof{\labof{\graph}(e')}}$. The proof
of~\cite[Theorem 5.2]{CourcelleV} builds a copyless \mso-transduction
scheme with the following parameters: \begin{compactitem}
\item $C$ is interpreted as the set of vertices of the input graph
  that become the vertices of the parse tree; these vertices
  correspond to a choice of internal vertices $c_p\in\vertof{\graph}$,
  for each rule $p=[u \rightarrow (\graph,f_1,\ldots,f_k)]$ that
  labels an edge of the parse tree,
\item $E_{i,p,j}$ is interpreted as the set of edges from the input
  graph that correspond to the $j$-th edge in the c-graph $\graph$
  from the productive rule $p$, enumerated in some fixed total order,
  that occurs, moreover, as the label of an $i$-edge of the parse
  tree; we denote by $\mathcal{E}$ the set of parameters $E_{i,p,j}$.
\end{compactitem}
Using the bijection $\gamma$ between the rules of graph grammars and
those of SIDs (Lemma \ref{lemma:canonical-hr}), one can build an
\mso-transduction scheme having the parameters $\mathcal{W} \isdef
\set{C} \cup \{E_{i,\gamma(p),j} \mid E_{i,p,j} \in
\mathcal{E}\}$. This transduction scheme is the composition of the
\mso-transduction scheme from the proof of ~\cite[Theorem
  5.2]{CourcelleV} with an \mso-transduction scheme that relabels the
edges of the parse trees by $\gamma^{-1}$. By Prop. \ref{prop:comp-trans}, the
composition of \mso-definable transductions is \mso-definable.

\begin{corollary}\label{cor:regular-sid-parsable}
  One can build a copyless \mso-transduction scheme
  $\scheme^\trparse_\asid=\langle\varphi^\trparse_\asid,\psi^\trparse_\asid,\\
  \set{\theta^\trparse_{\asid,\arule}}_{\arule\in\asid}\rangle$,
  with parameters $\mathcal{W}$, such that $\csem{\apred}{\asid}$ is
  $\trans$-parsable, where $\trans=\defd{\scheme^\trparse_\asid}$.
\end{corollary}
This result is used in the proof of the following statement:

\begin{lemma}\label{lemma:canonical-mso}
  $\csem{\apred}{\asid}$ is \mso-definable.
\end{lemma}
\proof{ It suffices to prove that $\parsetreeof{\asid}{\apred}$ is
  \mso-definable and use Corollary \ref{cor:regular-sid-parsable} and
  Theorem \ref{thm:bt} to infer that $\csem{\apred}{\asid}$ is
  \mso-definable. But this is a consequence of the fact that the rules
  of a regular SID produce the same parse trees as a regular graph
  grammar (modulo a one-to-one relabeling of edges) and that the set
  of parse trees is \mso-definable, by~\cite[Proposition 3.8 and
    Theorem 3.3]{CourcelleV}. \qed}

As a technical detail, the construction of the transduction scheme
from~\cite[Theorem 5.2]{CourcelleV} relies on a set of \mso\ formulae
$\chi_{i,p,w}(\mathcal{W},x,y)$, where $p = [u \rightarrow
  (\graph,f_1,\ldots,f_k)]$ is a productive grammar rule and $w \in
\vertof{\graph}$ is a vertex of the c-graph that describes the
hyperedge-replacement operation. Intuitively, these formul{\ae}
``define'' $w$ from the designated vertex $c_p$ of the copy of
$\graph$ introduced by the same $i$-edge of the parse tree. We use the
same idea and define similar formul{\ae} for regular SIDs:

\begin{lemma}\label{lemma:regular-sid-internal}
 For each productive rule $\arule\in\asid$ of the form
 (\ref{rule:prod}), with a distinguished existentially quantified
 variable $c_\arule\in\fv{\psi}$ and each variable (either
 existentially quantified or parameter) $z \in \fv{\psi}$ one can
 build an \mso\ formula $\chi_{i,\arule,z}(\mathcal{W},x,y)$ such
 that, for each c-graph $\graph\in\csem{\apred}{\asid}$, tree
 $\tree=\defdof{\scheme^\trparse_\asid}{\store}(\graph)$ for a store
 $\store$, where $\graph \models^{\overline\store} \charform{\tree}$,
 for a store $\overline\store$, that is canonical for
 $\charform{\tree}$, the following are equivalent: \begin{compactenum}
  \item $\graph \Models^{\store} \chi_{i,\arule,z}(\mathcal{W},x,y)$,
  \item $\store(x)=\overline\store(\atpos{c}{e}_\arule)$ and
    $\store(y)=\overline\store(\atpos{z}{e})$ for some $i$-edge
    $e\in\edgeof{\tree}$, such that $\labof{\tree}(e)=\arule$.
  \end{compactenum}
\end{lemma}
\proof{By the argument from the proof of~\cite[Lemma 5.6]{CourcelleV},
  using the one-to-one correspondence $\gamma$ between graph operations
  and SID rules (Lemma \ref{lemma:canonical-hr}). In particular,
  $\overline{\store}(\atpos{c}{e}_\arule)$ and
  $\overline{\store}(\atpos{z}{e})$ denote vertices introduced by the
  same edge $e$ of the parse tree $\tree$, see the definition of
  characteristic formul{\ae} (\ref{eq:charform}). \qed}

\subsection{\mso-Definable Rich Canonical Models}

We generalize the result of Lemma \ref{lemma:canonical-mso} to the set
$\rcsem{\apred}{\asid}$ of rich canonical $\asid$-models of
$\apred$. To this end, we build an \mso\ formula $\diseqform(x,y)$
that defines the endpoints of the edges labeled with $\diseq$, as in
Def. \ref{def:canonical-model} (\ref{it2:def:canonical-model}). This
definition uses the formul{\ae} $\chi_{i,\arule,z}$ (Lemma
\ref{lemma:regular-sid-internal}) and $\varphi^\trparse_\asid$
(Corollary \ref{cor:regular-sid-parsable}) and is given below:
\begin{align}
  \diseqform(x,y) \isdef \exists z_0 \exists \mathcal{W} ~.~
  \varphi^\trparse_\asid(\mathcal{W}) \wedge
  \hspace*{-4mm}\bigvee_{\begin{array}{c}
      \scriptstyle{\arule \in \asid},~ i \in \interv{1}{M} \\[-1mm]
      \scriptstyle{z_1 \neq z_2 \text{ occurs in } \arule}
      \end{array}} \chi_{i,\arule,z_1}(\mathcal{W},z_0,x) \wedge \chi_{i,\arule,z_2}(\mathcal{W},z_0,y)
\end{align}
where $M \isdef \max\set{\arityof{\arule} \mid \arule\in\asid}$.
\end{textAtEnd}

\subsection{Fission as an \mso-Definable Transduction}
\label{sec:fission}

Since $\sidsem{\apred}{\asid} =
\proj{\left(\transfusion(\rcsem{\apred}{\asid})\right)}{\alphabet}$
(Lemma \ref{lemma:fusion}) and $\rcsem{\apred}{\asid}$ is
\mso-definable for any regular SID $\asid$ (Lemma
\ref{lemma:rich-canonical-mso}), it would be sufficient to show that
$\transfission$ is an \mso-definable transduction to infer that
$\sidsem{\apred}{\asid}$ is \mso-definable (Theorem \ref{thm:bt}). As
discussed, this problem is open.

%

In this section, we address the simpler problem of defining
$\fission{k}$, for a constant $k\geq1$ that does not depend on the
input graph. We recall that $\fission{k}$ is the $k$-times composition
of the $\fission{1}$ operation that splits an arbitrary vertex $u$ in
two copies $(u,1)$ and $(u,2)$, redirecting all edges incident to $u$
to either $(u,1)$ or $(u,2)$, but not both. Hence, by
Prop. \ref{prop:comp-trans}, it is sufficient to define $\fission{1}$
by an \mso-transduction scheme.

Because each c-graph from $\fission{1}(\graph)$ has one more vertex than
$\graph$, the transduction scheme that defines $\fission{1}$ needs to
be $2$-copying, as having several input layers is the only way of
increasing the size of a definable transduction's output. We use a
parameter $X_1$ interpreted as a set containing the single vertex $u$
of the input c-graph that is split in two by fission. Further parameters
$\set{X^i_a}_{a\in\alphabet^\diseq,i\in\interv{1}{\arityof{a}}}$
contain edges of the input c-graph, labeled by $a$, that are incident to
$u$ on position $i$.  These parameters are used to redirect the edges
to the copies of $u$ from the first (if the edge belongs to the
interpretation of $X^i_a$) or second layer (otherwise). Formally, let
$\scheme^\trfission\isdef\tuple{\varphi,~ \set{\psi_i}_{i=1,2},~
  \set{\theta_{\overline{a}}}_{\overline{a}\in\alphabet^\diseq \otimes
    2}}$ be an \mso-transduction scheme, where:
\begin{align*}
\varphi \isdef & ~\mathtt{single}(X_1) ~\wedge~ \forall x .~ X_1(x) \rightarrow \mathtt{vert}(x) ~\wedge~
\forall x . \hspace*{-2mm}\bigwedge_{\begin{array}{c}
    \scriptscriptstyle{a\in\alphabet^\diseq} \\[-1mm]
    \scriptscriptstyle{i\in\interv{1}{\arityof{a}}}
\end{array}}\hspace*{-2mm}
X^i_a(x) \rightarrow \exists y ~.~ X_1(y) \wedge \mathtt{incid}^i_a(x,y)
\\[-3mm]
\psi_1 \isdef & ~\mathtt{true} \hspace*{2cm} \psi_2 \isdef ~X_1(x_1)
\end{align*}

\vspace*{-1.5\baselineskip}
\begin{align*}
\theta_{(a,1,i_2,\ldots,i_{\arityof{a}+1})} \isdef & ~ \edgrel{a}(x_1,\ldots,x_{\arityof{a}+1}) ~\wedge~
\hspace*{-6mm}\bigwedge_{\begin{array}{c}
    \scriptscriptstyle{a\in\alphabet^\diseq} \\[-1mm]
    \scriptscriptstyle{k\in\interv{2}{\arityof{a}+1}},~
    \scriptscriptstyle{i_k=1}
\end{array}}\hspace*{-4mm} X_1(x_k) \rightarrow X^{k-1}_a(x_1) ~\wedge~
\hspace*{-6mm}\bigwedge_{\begin{array}{c}
    \scriptscriptstyle{a\in\alphabet^\diseq} \\[-1mm]
    \scriptscriptstyle{k\in\interv{2}{\arityof{a}+1}},~
    \scriptscriptstyle{i_k=2}
\end{array}}\hspace*{-4mm} X_1(x_k) \wedge \neg X^{k-1}_a(x_1)
\\[-1mm]
\theta_{(a,2,i_2,\ldots,i_{\arityof{a}+1})} \isdef & ~\mathtt{false} \text{, for all } i_2, \ldots, i_{\arityof{a}+1} \in \set{1,2}
\end{align*}
The above definition uses the shorthands $\mathtt{single}(X)\isdef
\exists x.~ X(x) \wedge \forall y.~ X(y) \rightarrow y=x$,
$\mathtt{vert}(x)\isdef \bigvee_{a\in\alphabet^\diseq}\exists y_1
\ldots \exists y_{\arityof{a}}.~
\edgrel{a}(x,y_1,\ldots,y_{\arityof{a}})$ and
$\mathtt{incid}^i_a(x,y)\isdef\exists z_1 \ldots \exists
z_{\arityof{a}}.~ \edgrel{a}(x, \\ z_1,\ldots,z_{\arityof{a}}) \wedge
y=z_i$. We show that $\scheme^\trfission$ is a \mso-transduction
scheme, that correctly defines the fission relation $\fission{1}
\subseteq \graphsof{\alphabet^\diseq} \times
\graphsof{\alphabet^\diseq}$:

\begin{lemmaE}[][category=proofs]\label{lemma:fission}
  $\scheme^\trfission$ is a
  $(\alphabet^\diseq,\alphabet^\diseq)$-transduction scheme, such that
  $\defd{\scheme^\trfission}(\graph)=\fission{1}(\graph)$, for each
  simple graph $\graph\in\graphsof{\alphabet^\diseq}$.
\end{lemmaE}
\begin{proofE}
  We prove first that $\scheme^\trfission$ satisfies condition
  (\ref{eq:trans}) from the definition of transduction schemes. Let
  $(b,i,j_1,\ldots,j_{\arityof{b}}), (c,i,k_1,\ldots,k_{\arityof{c}})
  \in \alphabet^\diseq \otimes 2$ be tuples, such that:
  \[\graph \Models^\store \theta_{(b,i,j_1,\ldots,j_{\arityof{b}})}(x,y_1,\ldots,y_{\arityof{b}})
  \wedge
  \theta_{(c,i,k_1,\ldots,k_{\arityof{c}})}(x,z_1,\ldots,z_{\arityof{c}})\]
  for some c-graph $\graph\in\graphsof{\alphabet^\diseq}$ and some
  store $\store$. Then $\graph \Models^\store
  \edgrel{b}(x,y_1,\ldots,y_{\arityof{b}})$ and $\graph \Models^\store
  \edgrel{c}(x,z_1,\ldots,z_{\arityof{c}})$, i.e.,
  $\store(x)\in\edgeof{\graph}$, $\labof{\graph}(\store(x))=b=c$ and
  $\edgerelof{\graph}(\store(x))=\tuple{\store(y_1),\ldots,\store(y_{\arityof{b}})}
  = \tuple{\store(z_1),\ldots,\store(z_{\arityof{c}})}$. We are left
  with proving that $j_\ell=k_\ell$, for all
  $\ell\in\interv{1}{\arityof{b}}$. Let
  $\ell\in\interv{1}{\arityof{b}}$ and distinguish the following
  cases: \begin{compactitem}
  \item $j_\ell=1$ and $\store(y_\ell)\not\in\store(X_1)$: suppose, for a
    contradiction, that $k_\ell=2$. Then, $\store(z_\ell)\in\store(X_1)$,
    contradiction, thus $k_\ell=1$.
  \item $j_\ell=1$ and $\store(y_\ell)\in\store(X_1)$: in this case,
    $\store(x)\in\store(X_b^{\ell})$ and suppose, for a contradiction,
    that $k_\ell=2$. Then, $\store(x)\not\in\store(X^\ell_b)$,
    contradiction, thus $k_\ell=1$.
  \item $j_\ell=2$: in this case $\store(y_\ell)\in\store(X_1)$ and
    $\store(x)\not\in\store(X^\ell_b)$. Suppose, for a contradiction,
    that $k_\ell=1$. Since $\store(y_\ell)=\store(z_\ell)$, we have
    $\store(z_\ell)\in\store(X_1)$, thus $\store(x)\in\store(X^\ell_b)$,
    contradiction, leading to $k_\ell=2$.
  \end{compactitem}
  Let $\graph\in\graphsof{\alphabet^\diseq}$ be a simple graph and
  $\overline\graph \in \graph$ be a c-graph (recall that $\graph$ is
  the set of c-graphs isomorphic to $\overline\graph$). We prove that
  $\defd{\scheme^\trfission}(\graph)=\fission{1}(\graph)$ below, where
  both $\defd{\scheme^\trfission}(\graph)$ and $\fission{1}(\graph)$
  denote sets of graphs:

  \vspace*{.5\baselineskip}\noindent``$\subseteq$'' Let $H =
  \defdof{\scheme^\trfission}{\store}(\overline\graph)$ be the c-graph
  resulting from the transduction applied to $\overline\graph$, for a
  store $\store$. Then $\store(X_1)=\set{u} \subseteq
  \vertof{\overline\graph}$ consists of a single vertex, by the
  definition of $\varphi$. By the definitions of $\psi_1$ and
  $\psi_2$, we have $\vertof{H} = \vertof{\overline\graph} \times
  \set{1} \cup \set{(u,2)}$ and $\edgeof{H} = \edgeof{\overline\graph}
  \times \set{1}$. Let $\approx$ be the equivalence relation generated
  by the pair $\tuple{(u,1),(u,2)}$ and $h : \vertof{H_{/\approx}}
  \cup \edgeof{H_{/\approx}} \rightarrow \vertof{\overline\graph} \cup
  \edgeof{\overline\graph}$ be the mapping defined as
  $h(\set{(v,1)})=v$, for all $v \in \vertof{\overline\graph}
  \setminus \set{u}$, $h(\set{(u,1),(u,2)})=u$ and $h((e,1))=e$, for
  all $e \in \edgeof{\overline\graph}$. It is easy to check that $h$
  establishes an isomorphism between $H_{/\approx}$ and
  $\overline\graph$, hence $\overline\graph\in\fusion{1}(H)$, i.e., $H
  \in \fission{1}(\overline\graph)$. Because the choice of
  $\overline\graph$ was arbitrary and $\graph$ denotes the set of
  c-graphs isomorphic to $\overline\graph$, we obtain
  $\defd{\scheme^\trfission}(\graph) \subseteq \fission{1}(\graph)$.

  \vspace*{.5\baselineskip}\noindent''$\supseteq$'' Let $H \in
  \fission{1}(\overline\graph)$ be a c-graph. Then
  $\overline\graph\in\fusion{1}(H)$, i.e, there exist two distinct
  vertices $u,v\in\vertof{H}$ and an equivalence relation $\approx$
  generated by the pair $\tuple{u,v}$, that is compatible with $H$,
  such that $\overline\graph$ is isomorphic to $H_{/\approx}$ and let
  $h : \vertof{\graph} \cup \edgeof{\graph} \rightarrow
  \vertof{H_{/\approx}} \cup \edgeof{H_{/\approx}}$ be the bijection
  that establishes this isomorphism. Let $\store$ be a store such that
  $\store(X_1)=h(\set{u,v})$ and $\store(X^i_a)=\set{h(e) \mid e \in
    \edgeof{H},~ \edgerelof{H}(e)_i=u}$, for all $a \in
  \alphabet^\diseq$ and $i \in \interv{1}{\arityof{a}}$. Then
  $\overline\graph \Models^\store \varphi$, by the definition of
  $\varphi$. In particular, for each edge $h(e) \in \store(X^i_a)$, we
  have $\edgerelof{H}(e)_i = u$, which implies that
  $\edgerelof{\overline\graph}(h(e))_i = \set{u,v}$, by the definition
  of $\approx$ and $h$. Let $K=\defdof{\scheme^\trfission}{\store}$ be
  the c-graph defined by $\scheme^\trfission$ for the choice of
  parameters given by $\store$. We conclude by showing that $K$ and
  $H$ are isomorphic, for a bijection $g : \vertof{K}\cup\edgeof{K}
  \rightarrow \vertof{H}\cup\edgeof{H}$ defined as
  $g((h(\set{u,v}),1))\isdef u$, $g((h(\set{u,v}),2))\isdef v$,
  $g((h(\set{w}),1))\isdef w$, for all $w \in \vertof{H} \setminus
  \set{u,v}$ and $g((h(e),1))=e$, for all $e \in \edgeof{H}$.  Because
  the choice of $\overline\graph$ was arbitrary and $\graph$ denotes
  the set of c-graphs isomorphic to $\overline\graph$, we obtain
  $\fission{1}(\graph) \subseteq
  \defd{\scheme^\trfission}(\graph)$. \qed
\end{proofE}

\section{Rigidity}
\label{sec:rigidity}

We give a syntactic condition that guarantees the bounded
fusionability of the set of canonical models of a given formula.  This
provides us with everything needed to prove the main result (Theorem
\ref{thm:main}). As before, let $\asid$ be an all-satisfiable,
equality-free regular SID and $\apred$ be a nullary predicate.

The \emph{coloring} of an equality-free qpf-formula $\psi$ on the
alphabet of edge labels $\alphabet$ is the mapping $\funcolor{\psi} :
\fv{\psi} \rightarrow \pow{\alphabet}$, defined as $\funcolor{\psi}(x)
\isdef \set{a \in \alphabet \mid a(x, x, \ldots, x) \mbox{ occurs in }
  \psi}$, for all $x \in \fv{\psi}$. Intuitively, a fusion can only
join vertices that correspond to the store values of variables with
disjoint colors in a canonical model. For a productive rule $r$ of
$\asid$ and a $(\asid,\apred)$-parse tree $\tree$ denote by
$\nrof{r}{\tree}$ the number of edges labeled by $r$ in $\tree$.

A non-empty set $\asid^p \subseteq \asid$ of productive rules is
\emph{pumping} for $\asid$ and $\apred$ if the set $\set{ \min_{r \in
    \asid^p} \nrof{r}{\tree} ~|~ \tree \in
  \parsetreeof{\asid}{\apred}}$ is not finite. Intuitively, a pumping
set of rules can be unfolded any number of times to produce larger and
larger $\asid$-models of $\apred$.

\begin{definition}
A regular SID $\asid$ is \emph{rigid} for $\apred$ if, for any subset
$\asid^p \subseteq \asid$, that is pumping for $\asid$ and $\apred$,
any rules $\arule_1, \arule_2 \in \asid^p$, such that $\psi_i$ is the
largest qpf formula occuring in $\arule_i$ and $\vec{y}_i$ is the set
of existentially quantified variables in $\psi_i$, we
have $\funcol{\psi_1}(y_{1}) \cap \funcol{\psi_2}(y_{2}) \neq
\emptyset$, for all $y_i\in\vec{y}_i$ and $i=1,2$. 
\end{definition}
The above definition is effective as the pumping
sets of a given SID can be computed:

\begin{propositionE}[][category=proofs]\label{prop:pumping-sets}
  The set $\set{\asid^p \subseteq \asid \mid \asid^p \text{ is pumping
      for } \asid \text{ and } \apred}$ is effectively computable.
\end{propositionE}
\begin{proofE}
  Given an integer $m\geq1$, a set $L \subseteq \nat^m$ is
  \emph{linear} if it is of the form $\vec{b}_0 + \vec{g}_1 \nat + \ldots + \vec{g}_\ell
  \nat$ for some $\vec{b}_0 \in \nat^m$ and $\vec{g}_1,\ldots,\vec{g}_\ell \in \nat^m$,
  called respectively the basis and the generators of $L$.  A set $S
  \subseteq \nat^m$ is called \emph{semilinear} if it is a finite
  union of linear sets. The Parikh image of a word $w \in \Sigma^*$
  where $\Sigma = \set{\alpha_1,\alpha_2,\ldots,\alpha_m}$ is a finite
  alphabet, is the vector $pk(w) \isdef \tuple{n_1,n_2,\ldots,n_m}$
  from $\nat^m$ containing the number of occurrences $n_i$ of every
  symbol $\alpha_i$ in $w$. It is well-known that the Parikh image of
  a context free-language subset of $\Sigma^*$ is semilinear and that
  the construction of this semilinear set is effective from a
  context-free grammar defining that
  language~\cite{DBLP:journals/ipl/EsparzaGKL11}.

  First, it is straightforward to construct a context-free grammar
  whose language corresponds to infix linearizations of its parse
  trees. In this grammar, the terminals (that is, the alphabet) are
  the productive rules of $\asid$, the non-terminals are the
  predicates $\preds$ of $\asid$ and the productions are derived from
  the rules of $\asid$ as follows: \begin{compactitem}
  \item $\qpred \rightarrow r$, for a productive rule $r$ of the form
    \ref{it1:def:regular-sid} in Def.~\ref{def:regular-sid},
  \item $\qpred \rightarrow r \ppred_1 \ldots \ppred_k$, for a
    productive rule $r$ of the form \ref{it2:def:regular-sid} in
    Def.~\ref{def:regular-sid},
  \item $\ppred \rightarrow \ppred \qpred$, for a non-productive rule
    of the form \ref{it3:def:regular-sid} in
    Def.~\ref{def:regular-sid},
  \item $\ppred \rightarrow \qpred_1 \ldots \qpred_\ell$, for a
    non-productive rule of the form \ref{it4:def:regular-sid} in
    Def.~\ref{def:regular-sid}.
  \end{compactitem}
  Then, a set $\asid^p$ of productive rules is pumping for $\asid$ and
  $\apred$ if and only if the Parikh image of the language generated
  by the above grammar starting from non-terminal $\apred$ contains a
  linear set $L$ such that, for every $r \in \asid^p$ the set $L$ has
  a generator $\vec{g}$ whose value for the $r$ symbol is
  non-zero. \qed
\end{proofE}

The rigidity condition is sufficient for bounded fusionability, as the
following lemma shows. The dual is however, not true, as shown by the
counterexample in Fig. \ref{fig:sids} (a) which is bounded fusionable,
but not rigid.

\begin{lemmaE}[][category=proofs]\label{lemma:rigid-bf}
  If $\asid$ is rigid for $\apred$ then $\fbof{\rcsem{\apred}{\asid}}
  \leq B$, for an effectively computable $B \geq 0$.
\end{lemmaE}
\begin{proofE}
  Let $m\isdef\cardof{\asid}$ and $\bigcup_{j=1}^n (\vec{b}_j +
  \vec{g}_{j1}\nat + \ldots + \vec{g}_{jk_j}\nat) \subseteq \nat^m$ be
  the Parikh image of (infix linearizations) of the set of trees
  $\parsetreeof{\asid}{\apred}$, constructed as explained in the proof
  of Lemma~\ref{prop:pumping-sets}, where $\vec{b}_j,
  \vec{g}_{j1},\ldots \vec{g}_{jk_j} \in \nat^m$, for all $j\in
  \interv{1}{n}$.  For a vector $\vec{b} \isdef \tuple{b_1,\ldots b_m}
  \in \nat^m$ denote the $|\vec{b}| = \sum_{i=1}^m b_i$.  We define $B
  \isdef \max_{j=1}^n |\vec{b}_j| \cdot \max_{\arule \in \asid}
  \size{\arule}$, where $\size{\arule}$ is the size of the maximal qpf
  formula that occurs in $\arule$.

  It is an easy observation that, in any parse tree $\tree \in
  \parsetreeof{\asid}{\apred}$ there exists at most $\max_{j=1}^n
  |\vec{b}_j|$ productive rules which are not pumping.  Also, note
  that in every productive rule $\arule$ there exists at most
  $\max_{\arule \in \asid} \size{\arule}$ quantified variables
  occurring in the maximal qpf formula from $\arule$. Let $\graph \in
  \rcsem{\apred}{\asid}$ be a c-graph, $T \in
  \parsetreeof{\asid}{\apred}$ a parse tree and a store $\store$ on
  $\fv{\charform{\tree}}$ such that $\graph \models^{\store}
  \charform{\tree}$. Using the observation above, there exists at most
  $B$ vertices in $\graph$ which are the image of quantified variables
  from non-pumping rules.

  Assume by contradiction that $\fbof{\graph} > B$.  Then, there must
  exists some equivalence compatible with $G$ which is $k$-generated,
  for some $k > B$.  Then, observe that any equivalence relation
  compatible with some $\graph$ which is at least $(B+1)$-generated
  will necessarily pair two vertices which are not among the (at most)
  $B$ ones mentioned above, hence, they are necessarily the images by
  $\store$ of quantified variables occurring in pumping rules.  But,
  this contradicts the rigidity condition, ensuring that vertices
  corresponding to quantified variables from pumping rules in cannot
  be fused because of non-disjoint sets of colors. \qed

\end{proofE}

\noindent
For instance, we can ``rigidify'' the SIDs from Fig. \ref{fig:sids} by labeling all
but a fixed number of fusionable variables with the same relation
atom. For instance, we can change the rule $\bpred(x_1, x_2)
\leftarrow \exists y ~.~ b(x_1,x_2) * c(y,x_2) * \bpred(x_1, y)$ into
$\bpred(x_1, x_2) \leftarrow \exists y ~.~ b(x_1,x_1) * b(x_1,x_2) *
c(y,x_2) * \bpred(x_1, y)$ in Fig. \ref{fig:sids} (a) and rule
$\cpred(x_1) \leftarrow \exists y_1 \exists y_2 ~.~ b(x_1,y_1) *
c(y_1, y_2)$ into $\cpred(x_1) \leftarrow \exists y_1 \exists y_2 ~.~
b(x_1,y_1) * c(y_1, y_2) * b(y_1,y_1) * b(y_2,y_2)$ in
Fig. \ref{fig:sids} (b).

\section{Conclusions}

We tackle the problem of defining an expressive fragment of Separation
Logic of Relations that describes sets of graphs of bounded tree-width
and, moreover, has an effective equivalent \mso\ translation. This
fragment ensures the decidability of entailments. We leverage from
existing results on \mso-definable hyperedge replacement graph
grammars and \mso-definable transductions, applied to the sets of
models of a substructural logic with separating conjunction
interpreted by standard disjoint union of the relations defined by the
edge labels of a graph.

\bibliographystyle{splncs04}
\bibliography{refs}

\appendix
\section{Fission is not \msone-Definable}
\label{app:fission}

We denote by \msone\ the variant of \mso\ in which the atoms
$\edgrel{a}(x_1,\ldots,x_{\arityof{a}+1})$ are replaced by
$a(x_1,\ldots,x_{\arityof{a}})$, with the following semantics:
\[\graph \Models^\store a(x_1,\ldots,x_{\arityof{a}}) \iff
\text{there exists } e \in \edgeof{\graph} \text{, such that }
\labof{\graph}(e)=a \text{ and }
\edgerelof{\graph}(e)=\tuple{\store(x_1), \ldots,
  \store(x_{\arityof{a}})}\] It is easy to see that \msone\ is a
fragment of \mso\ because each atom $a(x_1,\ldots,x_{\arityof{a}})$
can be translated as $\exists y ~.~
\edgrel{a}(y,x_1,\ldots,x_{\arityof{a}})$, where $y$ does not occur
anywhere else in the context of the atom.

\begin{proposition}
  The transduction $\transfission$ is not \msone-definable.
\end{proposition}
\proof{
  Consider the following \msone\ formula written using the
  relation symbols $\erel$ binary and $\acrel$ unary:
  \begin{align*}
    \mathit{Hamiltonian} \isdef & \forall x \exists! y ~.~ \erel(y,x) \wedge \acrel(y) ~\wedge \\
    & \forall x ~.~ \acrel(x) \rightarrow \exists! y ~.~ [\erel(x,y) \wedge \acrel(y)] ~\wedge \\
    & \forall x ~.~ \neg\acrel(x) \rightarrow \neg \exists y ~.~ \erel(x,y) ~\wedge \\
    & \exists! x \forall y ~.~ \mathit{reach}_\erel(y,x)
  \end{align*}
  where $\exists! x . \psi(x)$ stands for $\exists x ~.~ \psi(x)
  \wedge \forall y ~.~ \psi(y) \rightarrow x = y$ and
  $\mathit{reach}_\erel(x,y)$ is the \msone\ formula stating that $y$
  is reachable from $x$ via a path that uses only pairs from the
  interpretation of $\erel$. This formula describes any structure
  consisting of a ring of $\acrel$-vertices, each attached to zero or
  more vertices not in the interpretation of $\acrel$. Note that the
  only possible pairs in a fusion consist of two $\neg\acrel$ or a
  $\neg\acrel$ and a $\acrel$ vertex. Moreover, by joining a
  $\neg\acrel$ with an $\acrel$ vertex, we obtain an $\acrel$-vertex.
  Let $\mathcal{H} \isdef \set{\graph \mid \graph \Models
    \mathit{Hamiltonian}}$ be the set of models of
  $\mathit{Hamiltonian}$. Then, the set of structures in which every
  vertex is labeled by $\acrel$ which, moreover, has a Hamiltonian
  $\erel$-cycle that visits every vertex exactly once, is the
  intersection of $\transfusion(\mathcal{H})$ with the set
  $\mathcal{C}\isdef\set{\graph \mid \graph \Models \forall x ~.~
    \acrel(x)}$, i.e., by fusion, every vertex becomes part of the
  cycle of $\acrel$-vertices.

  Suppose, for a contradiction, that $\transfission$ is an
  \msone-definable transduction. By Thm. \ref{thm:bt}, the set
  $\transfusion(\mathcal{H}) \cap \mathcal{C}$ is
  \msone-definable. However, this contradicts the fact that the set of
  Hamiltonian structures is not \msone-definable. To see this,
  consider the set $\mathcal{K} \isdef \set{K_{n,m} \mid n,m \geq 1}$
  of bipartite graphs with $n$ vertices on the left and $m$ on the
  right. Clearly, this set is defined by the \msone\ formula:
  \begin{align*}
    \mathit{Bipartite} \isdef & \exists X \exists Y ~.~ \neg\exists x ~.~ X(x) \wedge Y(x) ~\wedge \\
    & \forall x \forall y ~.~ X(x) \wedge \erel(x,y) \rightarrow Y(y) ~\wedge \\
    & \forall x \forall y ~.~ Y(x) \wedge \erel(x,y) \rightarrow X(y)
  \end{align*}
  Hence the set $\mathcal{H} \cap \mathcal{K} = \set{K_{n,n} \mid n
    \geq 1}$ is \mso-definable, contradicting the fact that
  \mso\ cannot express equality of sets. \qed}

\ifLongVersion
\else

\section{Proofs}
\label{app:proofs}
\printProofs[proofs]

\section{Material from Section \ref{sec:transductions}}
\label{app:transductions}
\printProofs[transductions]
\printProofs[richcanon]
\fi

\end{document}